\documentclass[manuscript,natbib,screen,nonacm]{acmart}

\usepackage[T1]{fontenc}
\usepackage[utf8]{inputenc}
\usepackage[american]{babel}
\usepackage[ruled,vlined,linesnumbered]{algorithm2e} %
\usepackage{mathtools}
\usepackage{bm}
\usepackage[all]{nowidow}
\usepackage{xcolor}
\usepackage{subcaption}
\usepackage{hyperref}
\usepackage{acronym}
\usepackage{enumitem}
\usepackage{multirow}
\usepackage{csquotes}
\usepackage[flushleft]{threeparttable}
\usepackage{siunitx}

\usepackage[capitalise]{cleveref}
\crefformat{section}{\S#2#1#3} %
\crefformat{subsection}{\S#2#1#3}
\crefformat{subsubsection}{\S#2#1#3}

\acmJournal{TOSEM}

\newcommand\free[0]{\vspace{1em}\noindent}

\definecolor{greylight}{RGB}{240,240,240}
\definecolor{greymedium}{RGB}{189,189,189}
\definecolor{greydark}{RGB}{99,99,99}

\usepackage{tcolorbox}
\tcbset{colback=greylight, colframe=greydark, arc=1pt, boxrule=0.75pt}
\newcommand{\summarybox}[2]{
	\begin{tcolorbox}
		\small{
			\textbf{#1:} #2
		}
	\end{tcolorbox}
}

\newacro{CI}{confidence interval}
\newacro{CI/CD}{Continuous Integration and Deployment}
\newacro{CV}{coefficient of variation}
\newacro{MAD}{median absolute deviation}
\newacro{QoS}{quality of service}
\newacro{RCIW}{relative confidence interval width}
\newacro{RMAD}{relative median absolute deviation}
\newacro{RMIT}{Randomized Multiple Interleaved Trials}
\newacro{SLA}{Service Level Agreement}
\newacro{SUT}{system under test}
\newacro{VM}{virtual machine}

\begin{document}

\author{Nils Japke}
\orcid{0000-0002-2412-4513}
\affiliation{%
    \institution{TU Berlin \& ECDF}
    \department{Scalable Software Systems Research Group}
    \city{Berlin}
    \country{Germany}}
\email{nj@3s.tu-berlin.de}

\author{Martin Grambow}
\orcid{0000-0001-6866-5461}
\affiliation{%
    \institution{TU Berlin \& ECDF}
    \department{Scalable Software Systems Research Group}
    \city{Berlin}
    \country{Germany}}
\email{mg@mcc.tu-berlin.de}

\author{Christoph Laaber}
\orcid{0000-0001-6817-331X}
\affiliation{%
    \institution{Simula Research Laboratory}
    \city{Oslo}
    \country{Norway}}
\email{laaber@simula.no}

\author{David Bermbach}
\orcid{0000-0002-7524-3256}
\affiliation{%
    \institution{TU Berlin \& ECDF}
    \department{Scalable Software Systems Research Group}
    \city{Berlin}
    \country{Germany}}
\email{db@3s.tu-berlin.de}

\title{µOpTime: Statically Reducing the Execution Time of Microbenchmark Suites Using Stability Metrics}



\begin{abstract}
    Performance regressions have a tremendous impact on the quality of software.
    One way to catch regressions before they reach production is executing performance tests before deployment, e.g., using microbenchmarks, which measure performance at subroutine level.
    In projects with many microbenchmarks, this may take several hours due to repeated execution to get accurate results, disqualifying them from frequent use in CI/CD pipelines.
    We propose µOpTime, a static approach to reduce the execution time of microbenchmark suites by configuring the number of repetitions for each microbenchmark.
    Based on the results of a full, previous microbenchmark suite run, µOpTime determines the minimal number of (measurement) repetitions with statistical stability metrics that still lead to accurate results.
    We evaluate µOpTime with an experimental study on 14 open-source projects written in two programming languages and five stability metrics.
    Our results show that (i) µOpTime reduces the total suite execution time (measurement phase) by up to 95.83\% (Go) and 94.17\% (Java), (ii) the choice of stability metric depends on the project and programming language, (iii) microbenchmark warmup phases have to be considered for Java projects (potentially leading to higher reductions), and (iv) µOpTime can be used to reliably detect performance regressions in CI/CD pipelines.
\end{abstract}

\maketitle

\textbf{ACM Reference format:}\\
Nils Japke, Martin Grambow, Christoph Laaber, and David Bermbach.\ 2025.\ {µOpTime}: Statically Reducing the Execution Time of Microbenchmark Suites Using Stability Metrics.\ \textit{ACM Trans.\ Softw.\ Eng.\ Methodol.}\ 34, 8, Article 231 (October 2025), 26 pages.\\
\url{https://doi.org/10.1145/3715322}

\section{Introduction}
\label{sec:introduction}

As modern software systems grow in complexity and size, they are prone to experiencing performance regressions.
To detect such regressions before deploying to production, benchmarking is an effective measurement-based solution.
Microbenchmarks are the benchmarking technique of choice on source code level,
similar to unit tests for functional testing,
which can be integrated into \ac{CI/CD} pipelines to automatically ensure performance requirements, ideally, after every code change.

Growing software projects often lead to growing microbenchmark suites, i.e., collections of individual microbenchmarks that are executed sequentially and repeatedly to evaluate a software's performance.
However, frequently executing large microbenchmark suites in \ac{CI/CD} pipelines is unrealistic because they often run for several hours or even days~\citep{GrambowUMBS, Grambow2021, laaber_dynamically_2020}.
One reason for these extensive execution times is the conservatively-configured number of repetitions for each microbenchmark, i.e., more repetitions generally lead to more reliable results.
Current standard practice is to configure repetitions across different levels (e.g., within a VM and across different VMs), where each level potentially adds noise to performance measurements (e.g., variability inherent to the microbenchmark, variability of cloud platform performance).
This leads to a high overall number of executions of each microbenchmark to guarantee sufficient result accuracy.
At the same time, executing microbenchmarks after they have reached a stable result is dispensable, as it does not improve the ability to detect performance regressions, and, consequently, wastes time and computational resources.
The state of practice, however, is using the same universal repetition configuration for all microbenchmarks, which provides an optimization opportunity to reduce execution time.

Previous research has focussed on reducing the execution time of microbenchmarks by executing the microbenchmarks until a statistical threshold is reached~\citep{georges_statistically_2007, he_statistics_2019}, dynamically reducing microbenchmark warmup phases~\citep{laaber_dynamically_2020,traini_ai-driven_2024}, and selecting subsets of microbenchmarks~\citep{de_oliveira_perphecy_2017, AlGhamdi2016, AlGhamdi2020, chen_perfjit_2020, GrambowUMBS, Grambow2021}.
However, none of these approaches distinguish between different repetition levels, only reduce the benchmark warmup phase and neglect the measurement phase, or do not execute potentially regression-exposing microbenchmarks at all.

In this paper, we propose µOpTime, an offline analysis approach which applies statistical methods to assess a microbenchmark's result stability and determine an improved execution configuration for each microbenchmark in a suite.
Our approach is the first to shorten the overall execution time of individual microbenchmarks across different repetition levels.
This leads to each microbenchmark having an individual execution time in each repetition level, as opposed to a universally set execution time in standard practice.
In contrast to existing approaches, which dynamically stop a benchmark during the execution when measurements are stable~\citep{he_statistics_2019, AlGhamdi2016, AlGhamdi2020, laaber_dynamically_2020, metior, traini_ai-driven_2024}, µOpTime statically creates an optimal measurement configuration based on measurements from a full suite execution once, which is then used in consecutive benchmark suite runs.
µOpTime is orthogonal to the existing approaches, and could potentially be combined, e.g., a performance engineer might first use µOpTime to reduce the execution time of all microbenchmarks, then during runtime reduce the set of microbenchmarks to be used~\cite{Grambow2021}, and finally dynamically optimize warmup time~\cite{laaber_dynamically_2020,traini_ai-driven_2024}.

Leveraging this, µOpTime (i) executes the full microbenchmark suite with a full execution configuration once, (ii) then determines the execution configuration with the shortest execution time, which leads to reliable results for each microbenchmark individually, by simulating smaller execution configurations and calculating for which configurations microbenchmarks are below a stability threshold, and (iii) thus collects a set of short-running configurations for subsequent code changes.

\free
To evaluate our approach, we address the following research questions:

\free
\textbf{RQ1: To which extent does µOpTime reduce the execution time of full microbenchmark suites without affecting their result accuracy?}

We apply our approach, as well as two baseline approaches (i.e., minimum and random repetition levels), to three Go projects and eleven Java projects with $4,304$ microbenchmarks in total.
Some Java projects originate from two different research data sets, which is why we introduce the terms data set 1 for the Go projects, and data set 2 and data set 3 for the two Java data sets.
Crucially, the benchmarking results in data set 2 do not use warmup iterations in their result data, so that we can study the influence later (cf.\ RQ3).
We observe that, compared to the baseline approaches, µOpTime provides better result accuracy, and handles the trade-off between execution time and accuracy better.
Using µOpTime, we can reduce the execution time by up to $95.83\%$ for Go (median of $76.8\%$), by up to $35.96\%$ for Java projects in data set 2 (median of $13.08\%$), and by up to $71.74\%$ for Java projects in data set 3 (median of $26.78\%$), where at least 80\% of the microbenchmarks being within 3\% of their original result with the full configuration.
Some configurations of µOpTime lead to higher time reductions, but especially for Java, this leads to lower accuracy (details in \Cref{sec:rq1}).
µOpTime shortened the duration from $57.5$ to $2.32$ minutes for \emph{zap}, the Go project with the largest absolute reduction, and from $178.19$ to $114.11$ hours for \emph{rxjava}, the Java project from data set 2 with the largest absolute reduction.
The Java project from data set 3 with the largest absolute time reduction was \emph{xodus} with a reduction of $35.05$ minutes.
The median execution time reductions were $33.6$ minutes for Go, $2.45$ hours for Java in data set 2, and $17.45$ minutes for Java in data set 3.
For some stability metrics, such as relative median absolute deviation, the result quality of Java projects is negatively impacted, though, as more than $50\%$ of microbenchmarks have results deviating more than $3\%$ from the baseline result.

\free
\textbf{RQ2: How do different stability metrics in µOpTime affect the execution time reduction?}

Using five different stability metrics to minimize execution configurations, we show using statistical tests that the choice of metric has no influence for Go projects and the relative median absolute deviation (RMAD) is most suitable.
For Java projects, however, bootstrapping-based methods such as relative confidence interval width (RCIW) are preferable, as the result quality is otherwise negatively impacted.
As an example, using RMAD for Java projects can lead to up to 64.1\% of microbenchmarks to deviate more than 10\% from their original result (\emph{rxjava}), while using an RCIW based on the median leads to at least 75.37\% of microbenchmarks to deviate \emph{less than} 1\%.

\free
\textbf{RQ3: How does the microbenchmark result accuracy change when considering the warmup phase of Java microbenchmarks?}

To study the effects of the Java Virtual Machine (JVM) warmup behavior on µOpTime, we use the Java projects in data set 2 and discard a fixed number of warmup iterations.
This can, counter-intuitively, \emph{decrease} the overall duration, as measurements afterwards are more stable than before, thus letting µOpTime optimize the execution configuration further (an overall decrease of up to $63.58\%$, median of $42.77\%$).
The Java project with the highest absolute reduction is again \emph{rxjava}, with the duration decreasing from $178.19$ to $81.12$ hours overall.
The median reduction in overall execution time is $8.75$ hours.
The result quality remains high, with all projects having more than $85\%$ of microbenchmarks within $3\%$ of the baseline result of the full configuration.

\free
\textbf{RQ4: How effective are the reduced execution configurations in detecting performance changes in CI/CD pipelines?}

To simulate a realistic CI/CD pipeline, we apply µOpTime to three Go and four Java projects (only data set 3) for successive versions and show that the reduced execution configurations detect the same performance changes as the full configurations.
For the three Go projects, the total execution times for all versions (38.75, 43.75, and 26.25 minutes) were reduced to 2.75, 1.75, and 1.1 minutes, while the execution times for the four Java projects (40.0, 78.0, 94.0, and 104.0 minutes) were reduced to 36.43, 66.93, 81.23, and 92.6 minutes, respectively.
The reduced configurations of µOpTime correctly detect the majority of the performance changes detected with the default configurations, experiencing only 3 and 29 false positives and 2 and 7 false negatives when executing 348 Go and 1,844 Java microbenchmarks, respectively.

\free
These results show that µOpTime significantly reduces the execution time of microbenchmark suites while keeping a result accuracy.
This closes the gap further towards enabling a continuous benchmarking with microbenchmark suites in \ac{CI/CD} pipelines~\cite{GrambowLehmannBermbach2019, waller_including_2015}.

\free
To summarize, the main contributions of this paper are:
\begin{itemize}
    \item µOpTime, an offline optimization approach based on statistical methods, which produces a minimal execution configuration for each microbenchmark, while ensuring the accuracy of the microbenchmark results.
	\item A microbenchmark experiment data set specifically collected to evaluate µOpTime in a realistic cloud environment, consisting of three open-source Go projects in five versions.
	\item An extensive experimental evaluation of µOpTime, including for regression testing in CI/CD pipelines, using our purpose-built data set and two publicly available data sets from prior research~\cite{laaber_dynamically_2020, laaber_dynamically_rp_2020, laaber_applying_2021, laaber_applying_2021_repl}.
	\item A \emph{cloud benchmark tool} for executing microbenchmark suites of arbitrary Go projects in cloud environments using \ac{RMIT} sequencing and a prototype implementation of µOpTime.
\end{itemize}

\section{Background}
\label{sec:background}
In this section, we give an overview of core concepts which the remainder of the paper relies on.

\nodotparagraph{Microbenchmarks} evaluate the performance of single subroutines (in the following: functions) on source code level by calling functions repeatedly with artificial input and gathering metrics, such as latency and memory usage.
We refer to the entire collection of microbenchmarks for a single software package as a \textit{microbenchmark suite}. 
During performance measurements, the entire suite is executed several times in succession (No.\ of \textit{suite runs}) and on multiple cloud instances to ensure repeatability (No.\ of \textit{instances}). 
In each run, every microbenchmark is executed several times successively (No.\ of \textit{iterations}), and each microbenchmark then calls the respective function as often as possible for a fixed period of time (benchmark \textit{duration}).
Together, these four parameters define an \textbf{execution configuration} for a microbenchmark suite.
Increasing these numbers yields more executions and thus more reliable benchmark results, however, the total execution time also increases accordingly. 

\begin{figure}[tbp]
	\centering
	\includegraphics[width=0.7\textwidth]{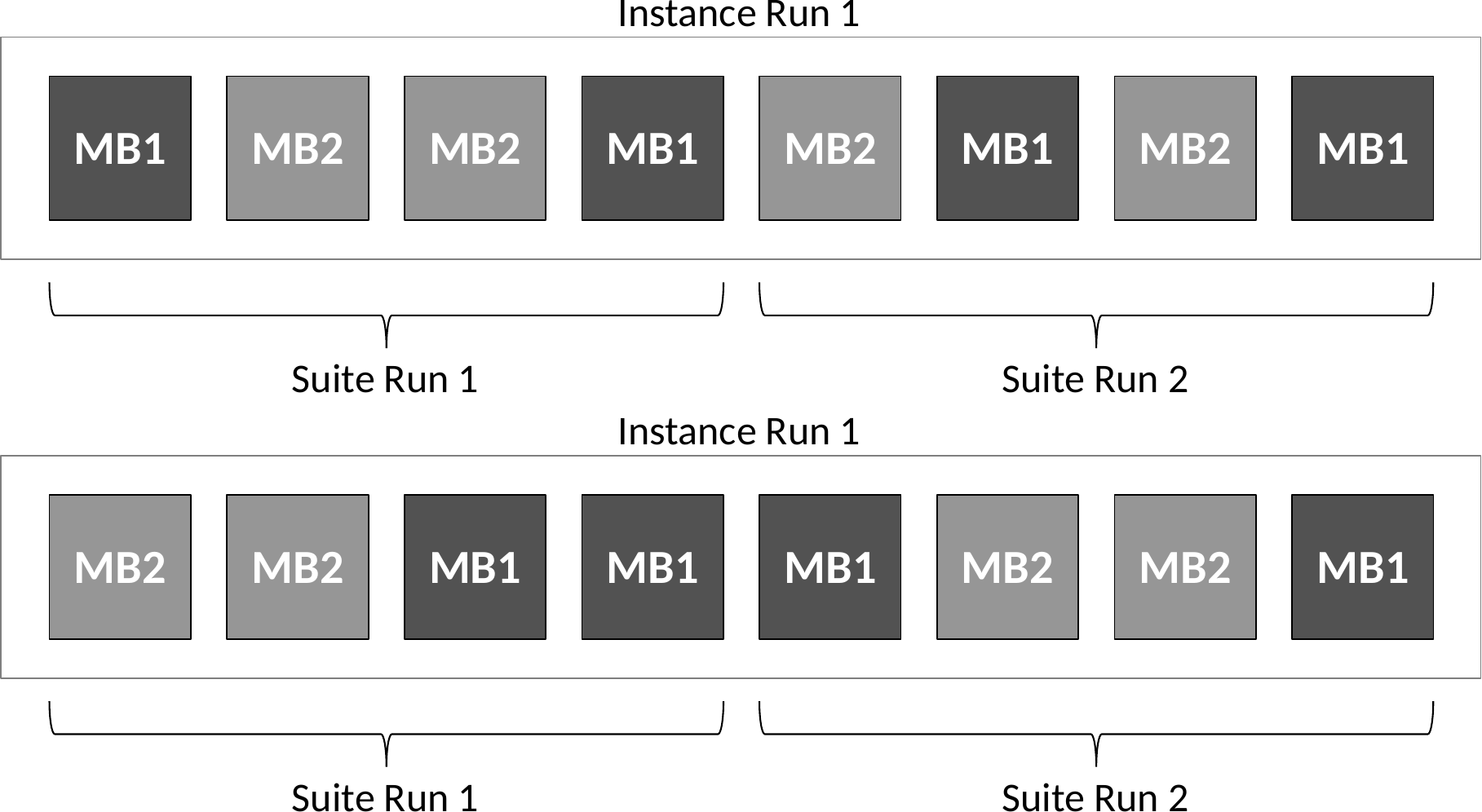}
  \caption{RMIT schema: Each microbenchmark is executed twice within each suite run, with each iteration shuffled to a random position inside the suite run. The full suite is run twice on two cloud instances concurrently.
  Not shown is the benchmark duration, which determines how long each iteration is executed.}
	\label{fig:rmit}
\end{figure}

\nodotparagraph{RMIT} (Randomized Multiple Interleaved Trials) is a randomization technique for eliminating and counteracting random fluctuation and measurement bias in microbenchmark results using multiple repetition levels~\cite{AbediOld,AbediCloud}.
Microbenchmarks in particular often run faster on repeated runs, because the underlying operating system and hardware use caching and speculative execution.
To get statistically sound, comparable results, microbenchmarks within a suite have to be shuffled to minimize influence from speculative execution and caching, since following iterations of a particular microbenchmark are not executed directly after one another.
In our setup, a suite is executed on a number of virtual cloud instances ($v$), the full suite is executed on each instance $s$ times, each benchmark within a suite is repeatedly executed for a number of iterations ($i$) following RMIT sequencing, and each iteration executes the evaluated function as often as possible for a fixed duration ($d$).
In total, we define a tuple $(v, s, i, d)$ as the execution configuration for a microbenchmark.
Each entry in this tuple also represents the different repetition levels.
It is possible to implement different repetition levels to counteract other sources of measurement fluctuations.
An example configuration can be seen in \Cref{fig:rmit}.
Using different repetition levels for microbenchmarks provides clear advantages for eliminating noise in measurements, but comes at the cost of increasing execution time and cost.

\nodotparagraph{Benchmarks in cloud environments} can be used to easily integrate performance evaluations in \ac{CI/CD} pipelines.
Here, fresh cloud \acp{VM} are provisioned for a benchmark evaluating the respective system under test's (SUT) performance.
Nevertheless, as cloud \acp{VM} vary in their performance~\cite{LaaberMBEval, GrambowLehmannBermbach2019, LaaberScheunerLeitner2019, LeitnerCito2016}, it is challenging to obtain reliable and repeatable results.
Thus, all benchmark experiments usually have to be repeated on multiple virtual cloud instances to ensure repeatability and relevance of results. 

\nodotparagraph{Stability Metrics} are standardized measures of variability, and we use them to determine the (in)stability of microbenchmark results.
We use standardized measures, i.e., a variability measure divided by a measure of location (e.g., the arithmetic mean), because they are dimensionless and therefore independent of scale.
Since the stability metrics are measures of relative variability, this means that a \emph{lower} stability metric points to better stability.
In particular, we apply:

\begin{itemize}
    \item The \textit{\ac{CV}} is a measure that standardizes the standard deviation of a probability distribution by dividing it by the arithmetic mean.
		\item The \textit{\ac{RMAD}} is a normalization of the \ac{MAD}, which is the median distance between the data and their median.
Here, the normalization is achieved by dividing the \ac{MAD} by the median of the data.
		\item The \textit{\ac{RCIW}} uses a normalization of a confidence interval width, i.e., the length of a \ac{CI} divided by a statistical average, to measure variability. 		
\end{itemize}

\nodotparagraph{Estimating Confidence Intervals} requires an estimator and an estimation method which constructs the confidence intervals around a point estimate. 

First, for the estimator, we choose the arithmetic mean or the median. 
Both are a kind of \enquote{average} that is of interest in performance engineering, and their \acp{CI} quantify their variability. 

Second, for the estimation method, we use two bootstrapping techniques: \emph{bootstrap percentile interval}~\cite{Hesterberg2015, Kalibera2020, EfronBook, DavisonBook} and \emph{bootstrap t-interval}~\cite{Hesterberg2015, EfronBook, DavisonBook}.
The former technique uses resampling to derive an empirical distribution for the estimator.
Then, the percentiles of that empirical distribution can be used to derive the \ac{CI}.
This technique has the advantage of being easy to implement, because it does not need a distributional assumption or other estimates.
It is also well-established within performance engineering~\cite{Kalibera2020, GrambowUMBS, laaber_predicting_2021}.
A disadvantage, however, is that percentile intervals tend to be too narrow for small sample sizes~\cite{Hesterberg2015}.
The latter technique generates an empirical t-distribution of the estimator using standardization (instead of using resampling to generate a distribution of estimates).
In order to standardize the estimates, we need to estimate the standard error during every generated sample.
For \ac{CI} based on the mean, we use the empirical standard deviation as an estimate for the standard error.
For \ac{CI} based on the median, however, we would need another technique based on bootstrap to estimate the standard error, leading to an \emph{iterated bootstrap}~\cite{EfronBook, DavisonBook}.
This makes the estimation of t-intervals more difficult to implement and their calculation more compute-intensive. 
Thus, we do not consider this estimator-method combination further.

\free
Relevant Metrics which we use in our optimizing approach are:
\begin{itemize}
    \item \textbf{CV}, normalized by the \emph{mean}
    \item \textbf{RMAD}, normalized by the \emph{median}
    \item \textbf{RCIW$_1$}, based on the \emph{mean} using a \emph{percentile interval}
    \item \textbf{RCIW$_2$}, based on the \emph{mean} using a \emph{t-interval}
		\item \textbf{RCIW$_3$}, based on the \emph{median} using a \emph{percentile interval}
\end{itemize}
For all our experiments, we estimate \acp{CI} using a confidence level of $99\%$ with $10,000$ bootstrap samples, which lines up with prior research~\cite{laaber_predicting_2021,Hesterberg2015}.

\section{Approach}
\label{sec:approach}

In this section, we introduce our approach µOpTime to statically reduce microbenchmark configurations, outline how µOpTime can be integrated into CI/CD pipelines, and introduce our prototype implementation.

µOpTime relies on an existing data set of microbenchmark results, e.g., from a previous full suite execution, which use a universal configuration of repetition levels for all microbenchmarks, and assesses the stability of each microbenchmark.
It then simulates all potential smaller configurations, which typically exhibit lower stability (i.e., they give a higher value for stability metrics) due to fewer performance measurements.
Since result variability decreases with increasing number of performance measurements, we can use this to balance the stability with execution time.
µOpTime calculates a chosen stability metric for each simulated configuration, and picks the smallest configuration for which the stability metric is below a chosen threshold, classifying it as \emph{stable}.
Both the stability metric and the threshold are input parameters of µOpTime.
The new individual configurations of repetition levels can then be used in future microbenchmarking runs to save time.
While changes to the underlying code may affect a microbenchmark's results, our approach assumes that it is rare that major changes, which fundamentally modify source code, occur in projects with a large code base~\cite{Alali2008, Leitner_Bezemer_2017}.

\subsection{Optimization}
\label{sec:optimization}

Our optimization algorithm works with data that use a universal configuration of repetition levels, such as the levels in an \ac{RMIT} setup, for all microbenchmarks.
These levels are represented by a tuple $(n_1, \dots, n_l) \in \mathbb{N}^+$ with $l \in \mathbb{N}^+$ being the number of levels chosen for the experiment, and each $n_i$ being the number of repetitions inside a level.
As an example, we might choose two levels for an experiment, where one level represents the \emph{iterations} ($i$) of a benchmark, and the other level represents their \emph{duration} ($d$).
We then have to choose the maximum number of repetitions for each level in the experiment, such as $i=3$ iterations and $d=5$ seconds duration, making the tuple represent the levels $(3, 5)$.\footnote{To simplify the use of durations of \ac{RMIT} setups here, we only consider full seconds of duration going forward.}
As a general rule, we order the tuple with the \emph{outermost} level first.
In our example, the outermost level is iterations (left), since each iteration runs for the duration (right).

\begin{algorithm}[tbp]
  \caption{µOpTime Optimization}\label{alg:optimization}
  \KwIn{\\
  $\bm{B}$: Set of microbenchmarks\\
  $\bm{e_{max}}$: Full execution configuration\\
  $\bm{M}$: Benchmarking results from full execution\\
  $\bm{sm}$: Stability metric\\
  $\bm{ts}$: Stability threshold value}
  \KwOut{\\
  $\bm{MIN}$: Set of minimal execution configurations}
  \Begin{
      $MIN \gets \emptyset$\\
      $E \gets findAllSmallerConfigurations(e_{max})$\\
      \ForAll{$b \in B$}{
          $S \gets \emptyset$\\
          $m \gets getMeasurements(e_{max}, M)$\\
          \ForAll{$e \in E$}{
              \If{$numberOfRepetitions(e) \geq 3$}{
                  $s_b \gets b$\\
                  $s_e \gets e$\\
                  $s_m \gets getMeasurements(e, M)$\\
                  $s_t \gets calculateExecutionDuration(e, M)$\\
                  $s_i \gets calculateStability(s_m, m, sm)$\\
                  $S \gets S \cup \{(s_b, s_e, s_m, s_t, s_i)\}$\\
              }
          }
          $S \gets filter(S, s_i \leq ts)$\\
          $S \gets findShortestDuration(S)$\\
          \If{$size(S) > 1$}{
              $S \gets findMinimumStability(S)$\\
          }
          $MIN \gets MIN \cup S$\\
      }
      \KwRet{$MIN$}
  }
\end{algorithm}

We note one possible execution configuration tuple $(n_1, \dots, n_l)$ as $e \in E$, where $E$ is the union of all possible execution configurations.
Moreover, $e_{max} \in E$ represents the full execution configuration, which is $(3, 5)$ in our example.

How the levels are chosen depends on the experiment being run, and which sources of randomness should be controlled for.
This will typically be chosen by the developers, in accordance to the needs of their software project (such as repeated runs across instances of the JVM, or repeated runs across different cloud VMs).
The goal of our optimization algorithm is to find a tuple $e \in E$ of repetitions such that the execution time of a microbenchmark is lower without significantly affecting its result quality.
We achieve this by using a stability metric that quantifies the variability of a microbenchmark result across repeated executions.
The idea is, that a microbenchmark of low(er) stability shows similar results with fewer measurements.

\Cref{alg:optimization} depicts our optimization algorithm.
As input, it requires (i) the set of microbenchmarks within a suite $B$, (ii) the full execution configuration tuple $e_{max}$, (iii) the measurement values from the full suite execution $M$, (iv) a stability metric $sm$, and (v) a stability threshold value $ts$ for classifying microbenchmark configurations as (in)stable.
(i)--(iii) are determined by the microbenchmarks, which a software project has implemented, and their settings for repetition levels.
(iv) and (v) are settings for µOpTime itself, and adjust the behavior.

The stability metric $sm$ (iv) measures the variability of individual microbenchmarks, while the stability threshold $ts$ (v) is used to determine, if the stability metric is low enough to consider the microbenchmark stable.
The key idea is that if we properly determine that results are stable, the result accuracy should not be affected much.
In all our later experiments, we set $ts = 0.01$, which means that the stability metric $sm$ needs to measure variability of $1\%$ or lower to consider a benchmark stable.
This value is informed by prior research on the variability of performance measurements.
Specifically, according to Georges et al.~\cite{georges_statistically_2007}, performance measurements often vary by approximately 3\%.
Huang et al.~\cite{huang_performance_2014} consider performance regressions between 3\% and 20\% as relevant.
We set $ts$ to be more conservative than these values in order for performance regressions of 3\% and over to still be detectable.

Initially, the set of minimal execution configurations is empty, and we determine all possible execution configurations based on the full configuration (lines 2 and 3).
Next, we initialize an empty set $S$ and look up the measurement values for the full execution configuration $m$ for all microbenchmarks, which will be our performance baseline to compare to (lines 4 to 6).
We then iterate over all levels of repetition starting with the smallest execution configurations up to the full one (lines 7 to 14).
This allows the algorithm to simulate benchmarking runs with fewer repetitions by only considering data that would have been collected in this scenario, up to the highest repetitions, where all measurement values are considered.
Continuing our example from above, we might start by considering only the first \emph{iteration}, of which we only use the first three seconds \emph{duration}, to simulate the levels of repetition with the tuple $(1, 3)$.
Since some statistical calculations, such as calculating the variance, do not work with too few data points, we use a lower limit of 3 repetitions per microbenchmark, i.e., the tuple $(1, 2)$ is skipped as we only have 2 measurements (line 8).
For all possible configurations $e \in E$ with more than three repetitions, we then collect the benchmark name and configuration, calculate the execution time for $e$, determine the stability based on the stability metric $sm$, and store all values in the set $S$ (lines 9 to 14).

Finally, we filter all entries in $S$, discard all with a larger stability than the given threshold $ts$, pick the entry with the shortest execution time, and add the minimal configuration for the respective benchmark to the result set $MIN$ (lines 15 to 19). 
If there are multiple with the same duration, we use the smaller stability metric as tie-break (lines 17 and 18).
After iterating over all microbenchmarks, $MIN$ contains the minimal individual execution configurations for each microbenchmark and is returned as result.

The time complexity of \Cref{alg:optimization} is $O(|B| \times |E|)$, where $|B|$ is the amount of microbenchmarks, and $|E|$ is the amount of all execution configurations smaller than $e_{max}$.
Since results of microbenchmarks do not depend on each other, the outer loop can be trivially parallelized.
For the Go projects we investigate later in our evaluation, $|B|$ is between 20 and 40 (though projects exist with $|B| > 1000$), while $|E| = 3 \times 5 \times 5$, as we use an \ac{RMIT} setup, where 3 repetition levels are optimized (see~\Cref{subsec:dataset}).

\subsection{Integrating µOpTime in CI/CD pipelines}
\begin{figure}[tbp]
	\centering
	\includegraphics[width=0.75\textwidth]{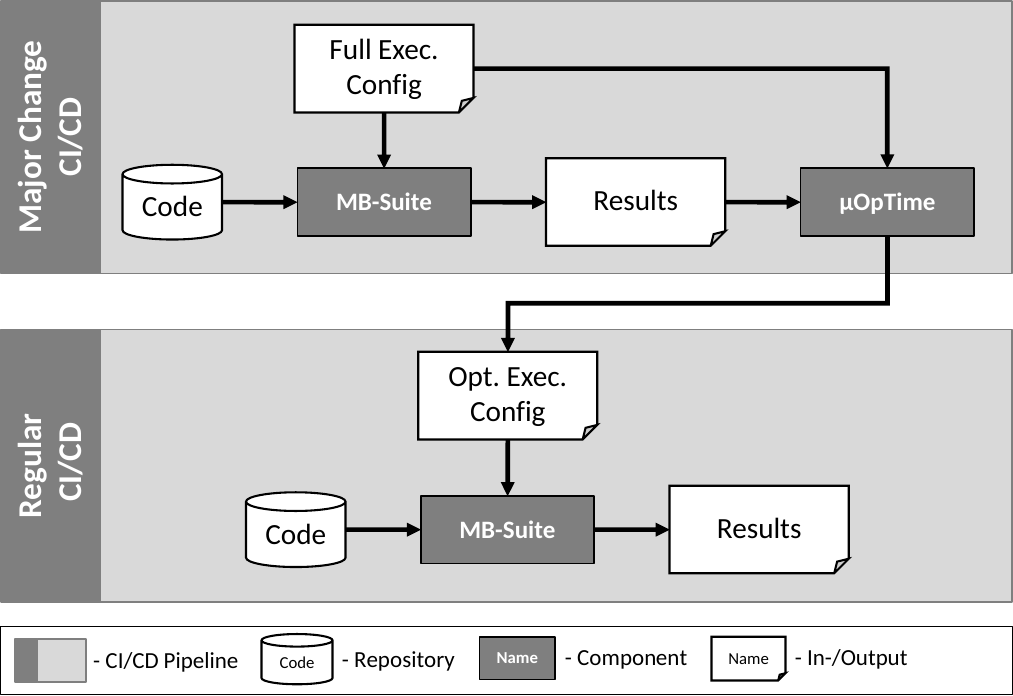}
    \caption{µOpTime in CI/CD pipelines: The optimization runs after every major change to identify minimal execution configurations for each microbenchmark individually.}
	\label{fig:setup}
\end{figure}

\Cref{fig:setup} shows the integration of µOpTime into CI/CD pipelines.
At first, because we do not know which microbenchmarks are more stable or unstable, they all have to be evaluated with the full configuration, which implies a long execution time.
µOpTime then uses these results, analyzes which microbenchmarks are stable, and identifies for each microbenchmark an individual execution configuration which ensures reliable results while avoiding unnecessary executions.
The individual and minimal execution configurations can then be used in regular CI/CD pipelines to quickly verify performance characteristics.
The minimal configurations must be re-determined regularly, depending on how often and how severely the software is modified, for those microbenchmarks affected by software modifications.

\subsection{Implementation}

µOpTime is implemented in an open-source Python prototype, available in our replication package~\cite{japke_uoptime_replication_package}.
It reads microbenchmarking results from a CSV file, identifies minimal execution configurations, and writes the results back into new CSV files.
All five stability metrics from \Cref{sec:background} (CV, RMAD, RCIW$_1$, RCIW$_2$, and RCIW$_3$) are already implemented, adding additional ones is simple.
The results contain the benchmarking score (mean or median performance depending on analyzed setup), \ac{CI}, and a lower bound on time saved (theoretical calculation, not including overhead) for both the full and reduced execution configurations.

\section{Experimental Evaluation}
To evaluate our approach, we use two existing public data sets and collect a third one, studying 14 open source projects written in Go (3 projects) or Java (11 projects) (details in~\Cref{subsec:dataset}).
We then explore our research questions RQ1 to RQ4 in \Cref{sec:rq1,sec:rq2,sec:rq3,sec:rq4}.

\subsection{Data Sets}
\label{subsec:dataset}

We use three different data sets to evaluate our approach.
The data sets we use contain microbenchmarking results in different programming languages (Go, Java), since we aim to provide generalizable results.
Furthermore, the two data sets containing Java microbenchmarking results are public data sets, and have been used in research before~\cite{laaber_dynamically_rp_2020, laaber_applying_2021_repl, laaber_dynamically_2020, laaber_applying_2021}.
Information on all data sets is summarized in \cref{tab:datasets}.

\begin{table}[tbp]
    \caption{Study Subjects:
    Data set 1 includes projects written in Go, and was collected for this study by us, while data set 2 and 3 include projects written in Java, and were collected by Laaber et al.\ for a study of optimizing warmup phases for microbenchmarks~\cite{laaber_dynamically_rp_2020, laaber_dynamically_2020} and a study on test case prioritization~\cite{laaber_applying_2021, laaber_applying_2021_repl}.
    Due to some gaps in the data of data set 3, we only use a subset of its software projects and versions.
    Data set 2 and 3 also have overlap in the projects they contain, giving us 14 Java project entries, with 11 distinct Java projects.}
    \label{tab:datasets}
    \centering
    \begin{threeparttable}
        \centering
        \begin{tabular}{clllcr}
            \toprule
            Data Set & Project\tnote{\textit{a}} & Symbol & Newest Ver.\tnote{\textit{b}} & \#~Ver. & \#~MB\tnote{\textit{c}}\\
            \midrule
            \multirow{3}{*}{\shortstack[c]{1\\[4pt]Go}} & \emph{prometheus/common} & G$_1$ & \texttt{296ec92} & 5 & 31\\
            & \emph{pelletier/go-toml} & G$_2$ & \texttt{9428417} & 5 & 35\\
            & \emph{uber-go/zap} & G$_3$ & \texttt{1ae5819} & 5 & 21 -- 46\\
            \midrule
            \multirow{10}{*}{\shortstack[c]{2\\[4pt]Java}} & \emph{raphw/byte-buddy} & J$_1$ & \texttt{c24319a} & 1 & 39\\
            & \emph{JCTools/JCTools} & J$_2$ & \texttt{19cbaae} & 1 & 148\\
            & \emph{jenetics/jenetics} & J$_3$ & \texttt{002f969} & 1 & 40\\
            & \emph{jmh-core-benchmarks}\tnote{\textit{d}} & J$_4$ & \texttt{a07e914} & 1 & 110\\
            & \emph{openjdk/jmh-jdk-microbenchmarks} & J$_5$ & \texttt{19cbaae} & 1 & 1,381\\
            & \emph{apache/logging-log4j2} & J$_6$ & \texttt{ac121e2} & 1 & 510\\
            & \emph{protostuff/protostuff} & J$_7$ & \texttt{2865bb4} & 1 & 31\\
            & \emph{ReactiveX/RxJava} & J$_8$ & \texttt{17a8eef} & 1 & 1,282\\
            & \emph{SquidPony/SquidLib} & J$_9$ & \texttt{055f041} & 1 & 367\\
            & \emph{openzipkin/zipkin} & J$_{10}$ & \texttt{43f633d} & 1 & 61\\
            \midrule
            \multirow{4}{*}{\shortstack[c]{3\\[4pt]Java}} & \emph{raphw/byte-buddy} & J$_{11}$ & \texttt{9aa0b2e} & 31 & 20 -- 39\\
            & \emph{jenetics/jenetics} & J$_{12}$ & \texttt{de28ba1} & 21 & 39 -- 58\\
            & \emph{JetBrains/xodus} & J$_{13}$ & \texttt{e2b902b} & 10 & 47 -- 74\\
            & \emph{openzipkin/zipkin} & J$_{14}$ & \texttt{36b7d32} & 20 & 34 -- 71\\
            \bottomrule
        \end{tabular}
        \begin{tablenotes}
            \small
            \item[\textit{a}] All projects are hosted on GitHub, except where noted.
            \item[\textit{b}] Only the newest version is listed. For commit hashes of all versions of data set 1, see our replication package~\cite{japke_uoptime_replication_package}, for data set 2 and 3, see the respective replication package~\cite{laaber_dynamically_rp_2020, laaber_applying_2021_repl}.
                Other versions than the newest one are only used for analyzing performance changes in RQ4 (\Cref{sec:rq4}).
            \item[\textit{c}] Whenever the amount of microbenchmarks changes between versions we used, the range is indicated.
            \item[\textit{d}] Module directory in repository: \url{https://github.com/openjdk/jmh}
        \end{tablenotes}
    \end{threeparttable}
\end{table}

The first data set contains microbenchmarking results of five versions of three different software projects written in Go.
We collected this data set specifically to evaluate µOpTime, as we needed cloud benchmarking data following an \ac{RMIT} execution configuration according to best practices using a different programming language than the public data sets.
We used an execution configuration of $(3, 3, 5, 5)$ when collecting the data to serve as a baseline.
Here, the first number represents \emph{instance runs}, the second number represents \emph{suite runs}, the third number represents \emph{iterations}, and finally the fourth number represents \emph{seconds per iteration}.
Grambow et al.~\cite{GrambowUMBS} have used a similar setup before.
Here, we increase the duration to 5 seconds, so that we can use another dimension, which affects stability, as this is effectively another way to increase repetitions of microbenchmarks.
We do this to ensure that the microbenchmarks run long enough for most of them to become stable, which is especially important in a cloud environment.
When using µOpTime on this data set, we do not apply it to the level of cloud instances, as that level removes a factor of variability not inherent to the microbenchmarks (cloud platform variance), so reducing this level would violate best practices, giving us effectively $(3, 5, 5)$ as the execution configuration for µOpTime.
As multiple cloud instances can be used in parallel, the execution time of this configuration is unaffected by the number of cloud instances, and becomes $3 \times 5 \times 5 = 75$ seconds per microbenchmark.
All data were collected on GCloud \texttt{n2-standard-2} instances (2 vCPUs, 8 GB memory) running Arch Linux.
We chose this particular instance type, as it is not affected by CPU bursting, as this affects software performance and introduces measurement bias.

The second and third data set used in our study were collected by Laaber et al.\ and contain microbenchmarking results from ten different software projects each, which were written in Java~\cite{laaber_dynamically_rp_2020, laaber_applying_2021_repl}.
As such, they use the \emph{Java Microbenchmark Harness} (JMH) to execute microbenchmarks.
Since the Java Virtual Machine (JVM) introduces further sources of variability by dynamically optimizing running code, it is best practice to discard the first few iterations as warmup iterations, and re-run experiments on different instances of the JVM called \emph{forks}.
While forks are not completely equivalent to our notion of a suite run, they are similar, as they also repeat the execution of the entire microbenchmark suite, so we will use forks in their place for this data set.
Both data sets were collected on bare-metal machines to minimize cloud influence on the measurements, unlike data set 1, for which we specifically used a cloud environment.

Laaber et al.\ originally used the second data set to study, how warmup phases typically used in microbenchmarking of Java software can be shortened~\cite{laaber_dynamically_2020}.
To that end, all data were collected without using JMH warmup iterations, so that each iteration is contained in the data set for further analysis.
Laaber et al.\ used an execution configuration of $(1, 5, 100, 1)$, since all experiments were run on a single bare-metal instance.
The JMH baseline for warmup iterations used by Laaber et al.\ consists of 5 iterations that last 10 seconds each before measurement begins, which corresponds to 50 iterations that last 1 second each in their execution configuration.

Similarly, Laaber et al.\ used the third data set to study test case prioritization (TCP) applied to microbenchmarks~\cite{laaber_applying_2021}.
This data set contains multiple versions for each software project, which were all executed using a configuration of  $(1, 3, 20, 1)$ with 20 warmup iterations of 1 second, which are already removed from the data.
Unfortunately, a few of the software projects measurement data had missing gaps, such as one fork not including one benchmark.
We decided to exclude every software project, which had some missing data, from further analysis, leaving us only with a subset of versions of four software projects for this data set.

\subsection{RQ1: To which extent can the execution time of full microbenchmark suites be shortened without affecting the accuracy of results?}
\label{sec:rq1}

\begin{figure*}[tbp]
	\centering

    \begin{subfigure}[b]{0.33\textwidth}
        \centering
        \includegraphics[width=\textwidth]{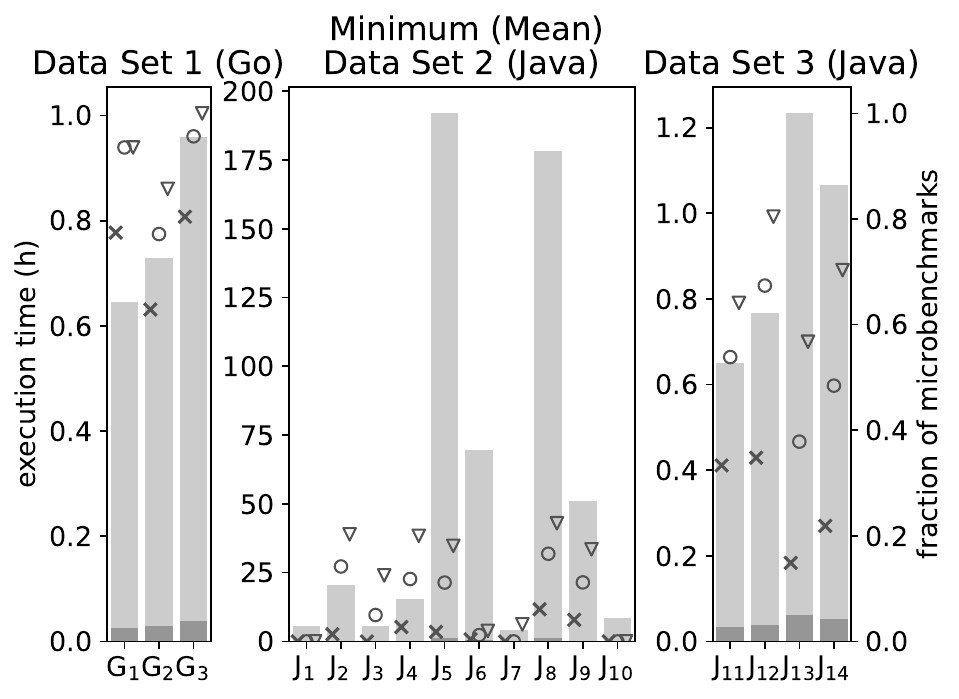}
        \caption{Results for the minimum baseline based on the mean.}
        \label{fig:rq1-rq2-time-savings-min-mean}
    \end{subfigure}
    \begin{subfigure}[b]{0.33\textwidth}
        \centering
        \includegraphics[width=\textwidth]{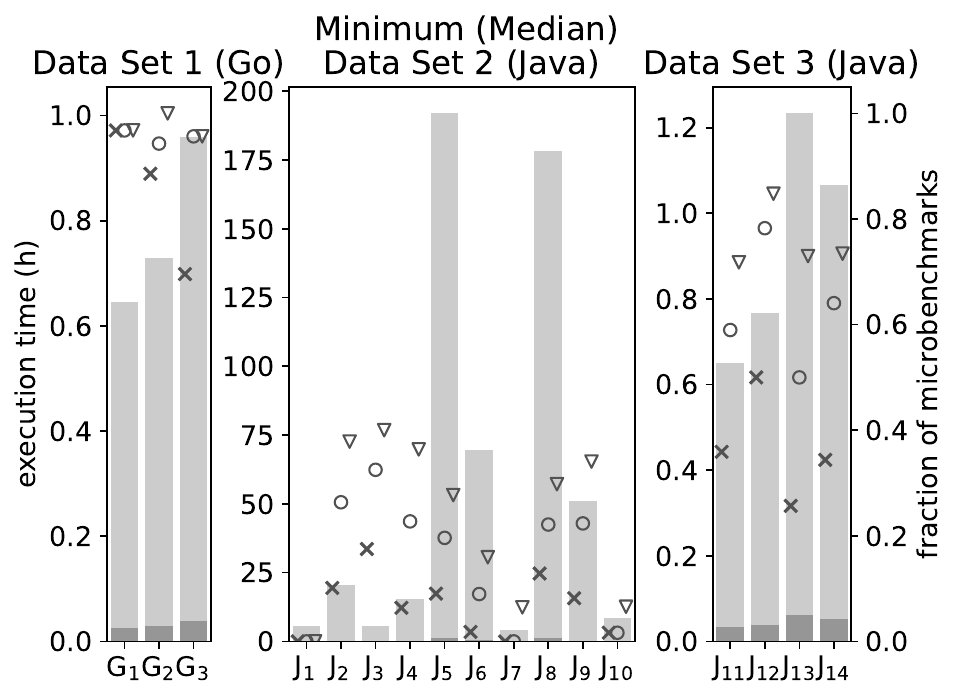}
        \caption{Results for the minimum baseline based on the median.}
        \label{fig:rq1-rq2-time-savings-min-median}
    \end{subfigure}
    \begin{subfigure}[b]{0.33\textwidth}
        \centering
        \includegraphics[width=\textwidth]{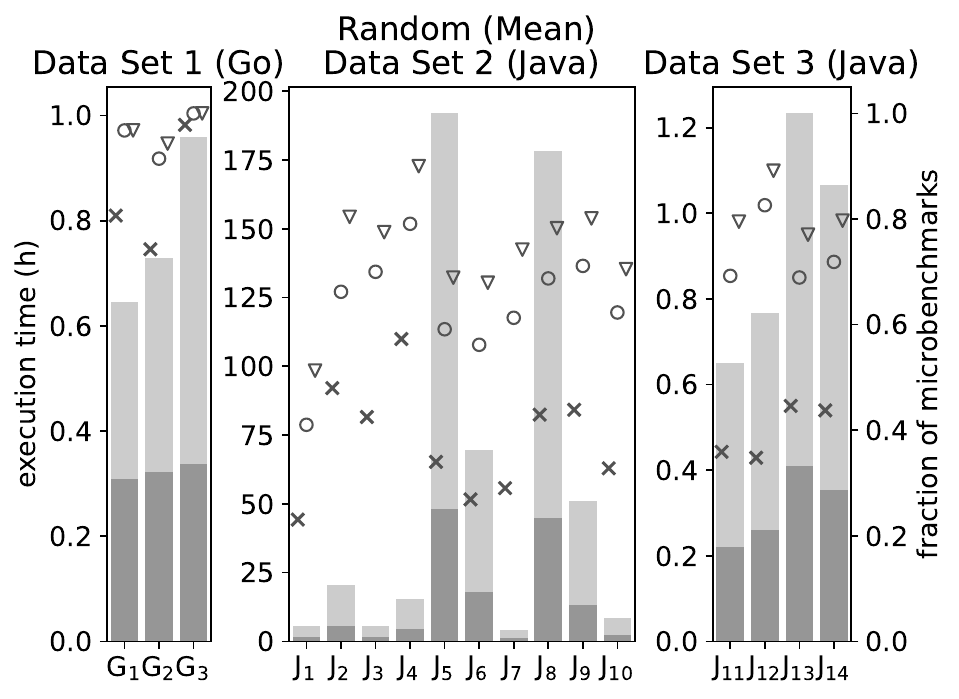}
        \caption{Results for the random baseline based on the mean.}
        \label{fig:rq1-rq2-time-savings-rand-mean}
    \end{subfigure}

    \begin{subfigure}[b]{0.33\textwidth}
        \centering
        \includegraphics[width=\textwidth]{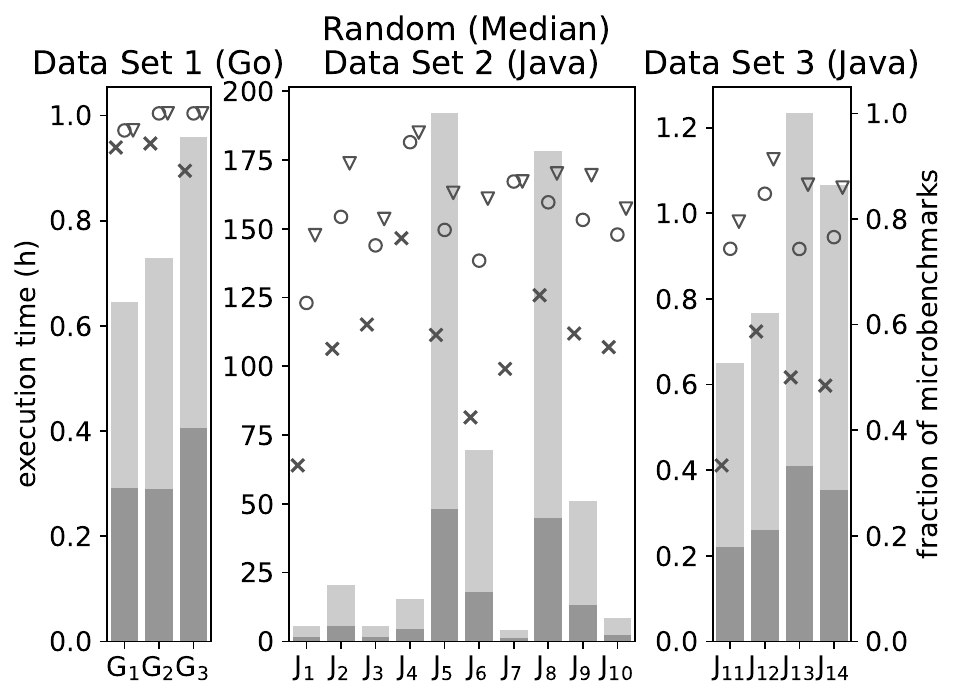}
        \caption{Results for the random baseline based on the median.}
        \label{fig:rq1-rq2-time-savings-rand-median}
    \end{subfigure}
    \begin{subfigure}[b]{0.33\textwidth}
        \centering
        \includegraphics[width=\textwidth]{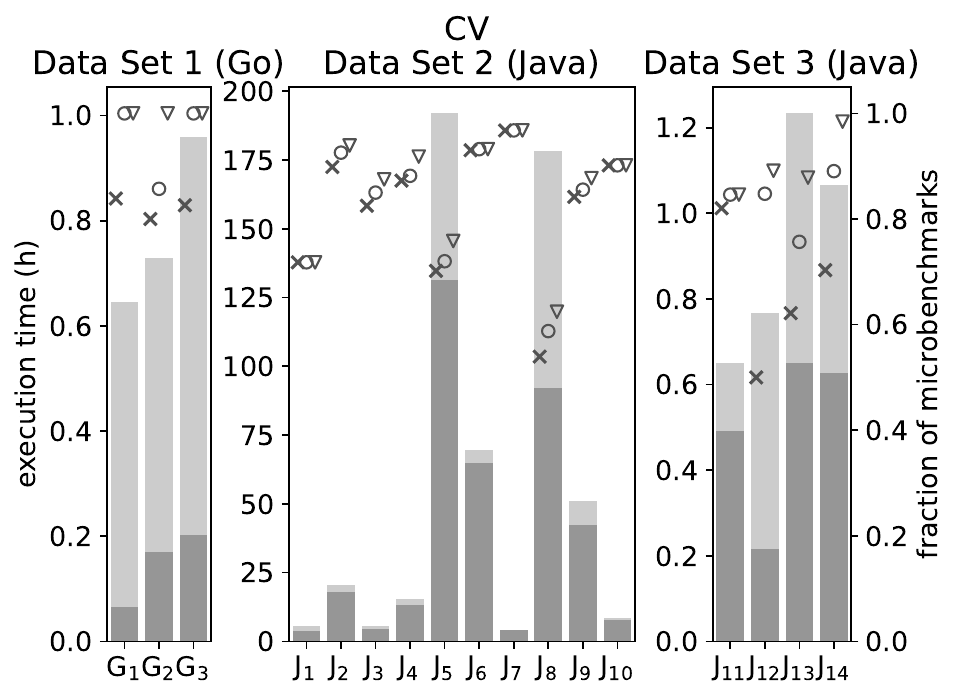}
        \caption{Results for the CV based on the mean.\\\phantom{a}}
        \label{fig:rq1-rq2-time-savings-cv}
    \end{subfigure}
    \begin{subfigure}[b]{0.33\textwidth}
        \centering
        \includegraphics[width=\textwidth]{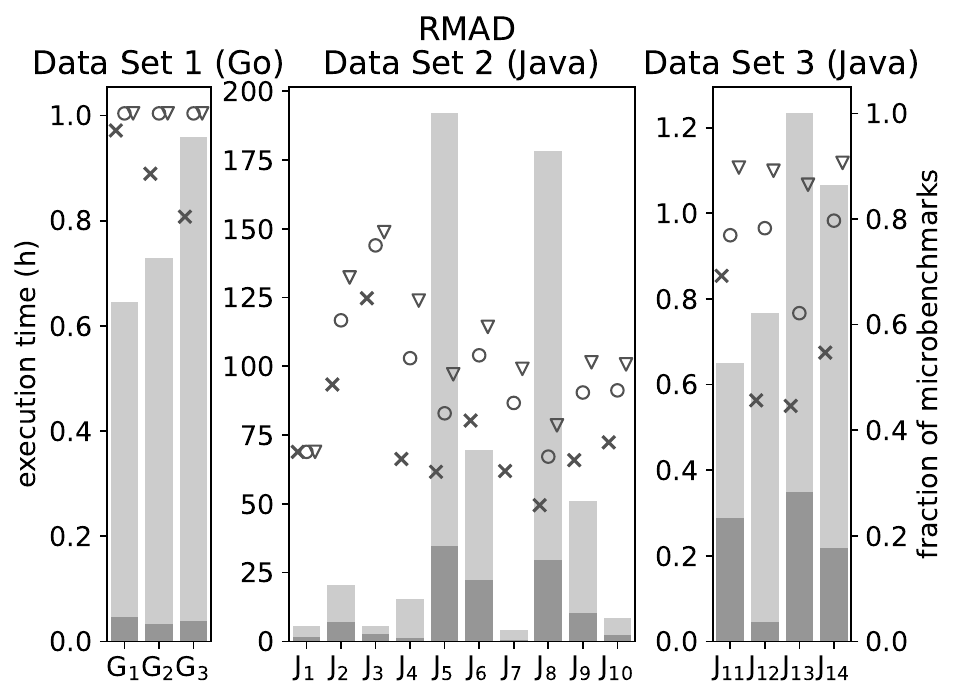}
        \caption{Results for the RMAD based on the median.}
        \label{fig:rq1-rq2-time-savings-rmad}
    \end{subfigure}

    \begin{subfigure}[b]{0.33\textwidth}
        \centering
        \includegraphics[width=\textwidth]{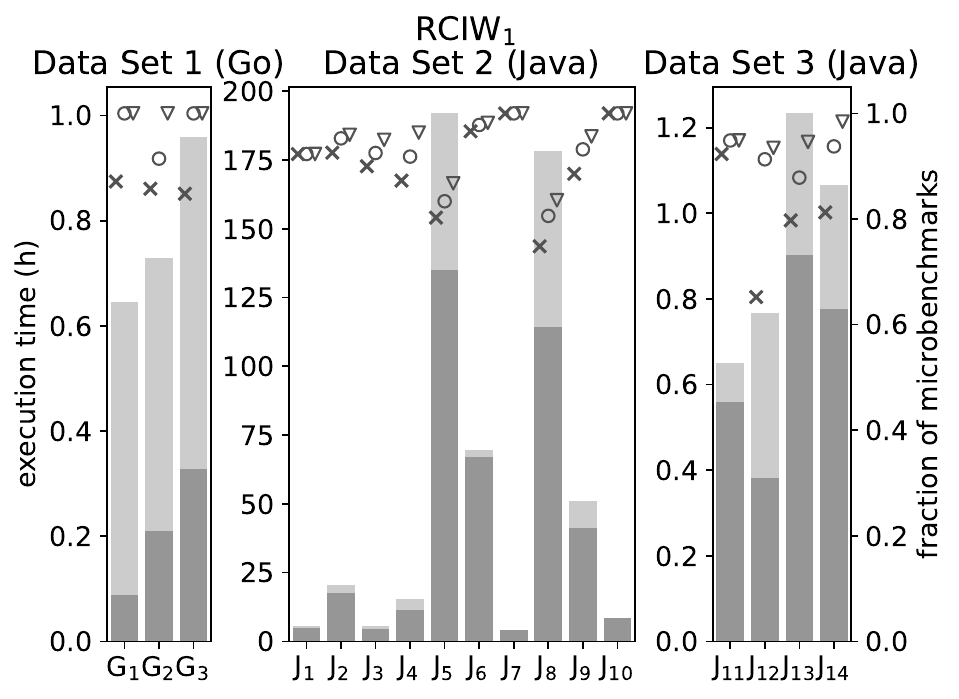}
        \caption{Results for the RCIW$_1$ based on the mean.}
        \label{fig:rq1-rq2-time-savings-rciw1}
    \end{subfigure}
    \begin{subfigure}[b]{0.33\textwidth}
        \centering
        \includegraphics[width=\textwidth]{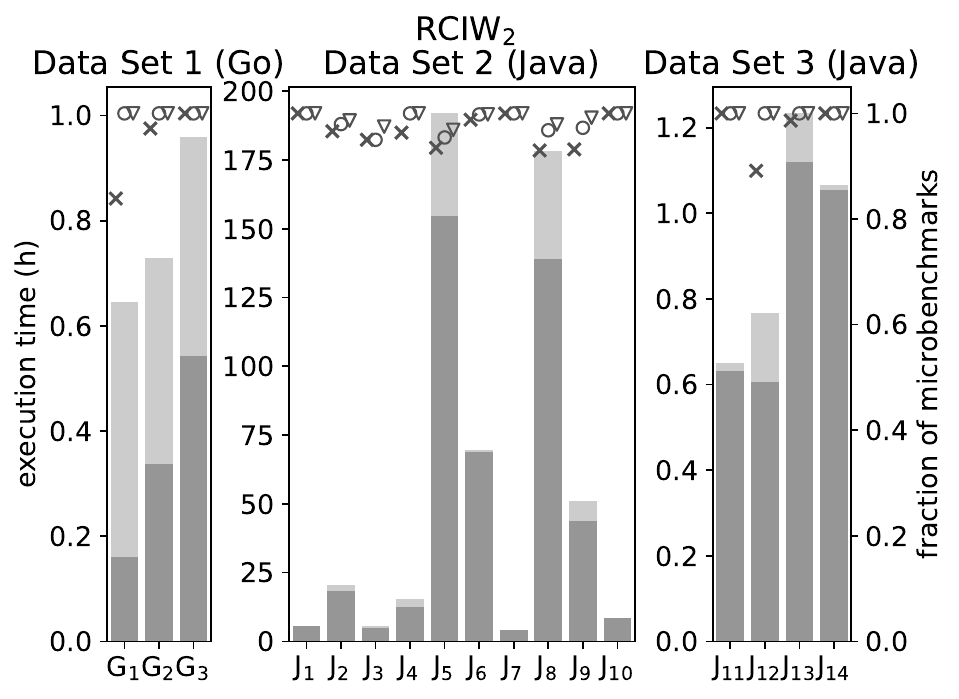}
        \caption{Results for the RCIW$_2$ based on the mean.}
        \label{fig:rq1-rq2-time-savings-rciw2}
    \end{subfigure}
    \begin{subfigure}[b]{0.33\textwidth}
        \centering
        \includegraphics[width=\textwidth]{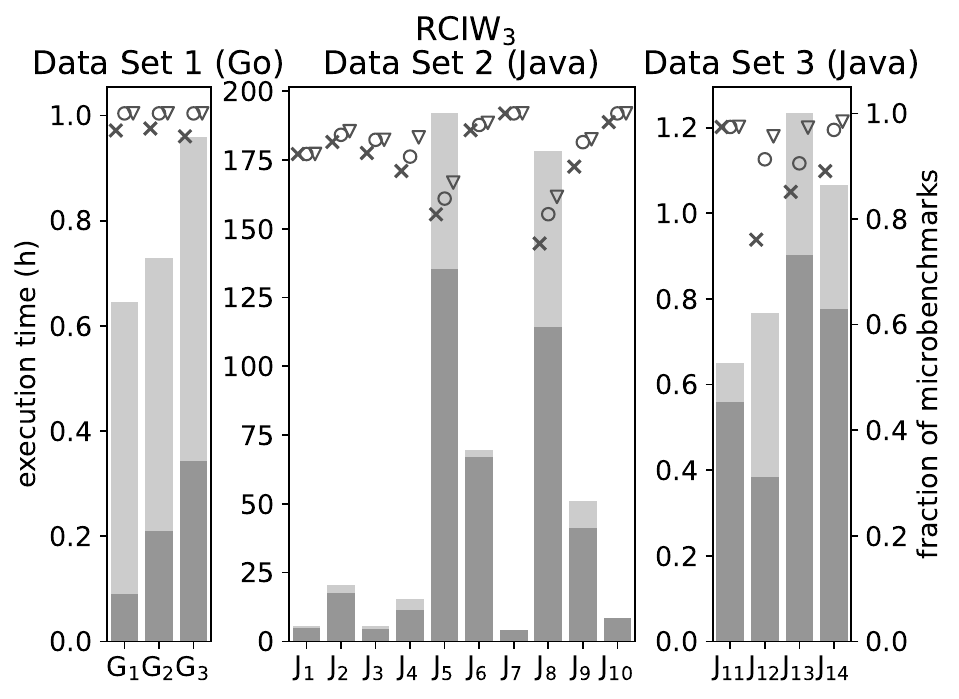}
        \caption{Results for the RCIW$_3$ based on the median.}
        \label{fig:rq1-rq2-time-savings-rciw3}
    \end{subfigure}
    
    \begin{subfigure}[b]{0.33\textwidth}
        \centering
        \vspace{1em}
        \includegraphics[width=0.8\textwidth]{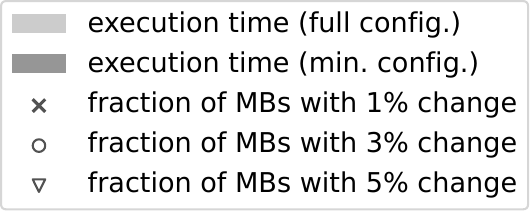}
    \end{subfigure}

    \caption{Time savings of µOpTime: The x-axis shows every project from data set 1, 2, and 3.
    The left y-axes of all three facets show the execution time in hours of the bars.
    Since the projects in each data set have a different number of microbenchmarks as well as a different execution configuration, the overall execution time varies strongly across data sets.
    Therefore, the left y-axes are scaled differently.
    The right y-axis shows different fractions of microbenchmarks (MBs) with low change rates after using µOpTime, represented by marks in the figure (e.g., the fraction of MBs with a change rate smaller than 1\%).}
	\label{fig:rq1-rq2-time-savings-all}
\end{figure*}

\paragraph{Context}
The first research question concerns the time savings of our approach.
We apply µOpTime to all study subjects from all three data sets to get a minimal configuration for each microbenchmark.
Then, we calculate the lowest possible time the minimal and full execution configurations take (i.e.,\ the execution time without any overhead), and compare how much time the minimal configuration saves.
For comparison, we include two simple baseline approaches, as other state-of-the-art approaches are orthogonal to µOpTime~(\Cref{sec:introduction}).
The first baseline approach (minimum baseline) always picks the smallest possible execution configuration, thereby maximizing time savings, while the second baseline approach (random baseline) picks a random execution configuration between the full and smallest execution configuration.

\paragraph{Methodology}
To assess the effects on accuracy, we calculate a \emph{change rate} $c$ for the performance results of each microbenchmark based on the results of the minimal and the full configuration ($r_{min}$ and $r_{full}$, respectively).
The change rate is defined in \Cref{eqn:changerate}, and quantifies a relative change in measured results when using the minimal instead of the full configuration.

\begin{equation}
    \label{eqn:changerate}
    c = \frac{|r_{min} - r_{full}|}{r_{full}}
\end{equation}

The results of a microbenchmark are either given as the mean, or the median of all measurements for that microbenchmark.
In practice, the mean is often used as a default (e.g., in the \emph{go test} tooling), while the median is chosen in some contexts for being more robust.
We use the mean for performance results, when the stability metric is based on the mean (\ac{CV}, \ac{RCIW}$_1$, \ac{RCIW}$_2$), and use the median for performance results, when the stability metric is based on the median (\ac{RMAD}, \ac{RCIW}$_3$).
For the baseline approaches, we include both results based on the mean and the median.

\paragraph{Results}
\Cref{fig:rq1-rq2-time-savings-all} shows the results for all individual study subjects.
The bars show a comparison between the execution time of the full and minimal configuration (with the time decreasing from the full bar to the smaller portion).
The different marks show the fraction of microbenchmarks with a change rate in performance results of 1\%, 3\%, and 5\%.
We choose these different thresholds to better understand, where substantial differences in the results may be able to obfuscate potential performance regressions.
This is informed by prior research, similar to how we set $ts$ (see \Cref{sec:optimization}), according to which performance measurements often vary by 3\%~\cite{georges_statistically_2007}, with performance regressions of 3\% to 20\% being relevant to practitioners~\cite{huang_performance_2014}.
Thus, if a high number of microbenchmarks has change rates lower than our thresholds, the majority of relevant performance changes can still be detected.

The minimum baseline (\Cref{fig:rq1-rq2-time-savings-min-mean,fig:rq1-rq2-time-savings-min-median}) shows the highest possible time savings (95\% to 99\%) but has the lowest accuracy.
It still fares well on Data set 1 (Go), with all projects having at least 62\% of their microbenchmarks within 1\% of the original result when comparing the mean, with that number increasing to 95\% when comparing the median.
For data set 2 (Java), the accuracy of the minimum baseline plummets, however, with all projects having at most 20\% of their microbenchmarks with a 5\% change or lower when comparing the mean.
Using the median gives slightly better results, where at most 40\% of each project's microbenchmarks are within a 5\% change of the original result.
The result quality for data set 3 (Java) is between data set 1 and 2 for the minimum baseline, with all projects having at least 50\% of their microbenchmarks with a 3\% change or lower.

The random baseline (\Cref{fig:rq1-rq2-time-savings-rand-mean,fig:rq1-rq2-time-savings-rand-median}) shows time savings between 50\% to 60\% for data set 1 (Go), 70\% to 75\% for data set 2 (Java), and around 66\% for data set 3 (Java).
The accuracy loss is miniscule for data set 1, with the worst performing project still having 74.3\% of microbenchmarks with a less than 1\% change when comparing the mean, which jumps to 89.1\% when comparing the median.
For the java projects, the accuracy loss slightly higher.
For data set 2 (java), at least 41\% of each project's microbenchmarks have a 3\% change or lower when comparing the mean, with that number increasing to 64.1\% when comparing the median.
The results on data set 3 (Java) are better than on data set 2, with at least 68.9\% of each project's microbenchmarks showing a 3\% change or lower when comparing the mean, with that number increasing to 74.3\% when comparing the median.

Out of the stability metrics, \ac{RMAD} has the highest time savings, reducing the execution time by 92.9\% to 95.97\% (Data set 1, Go), 49.66\% to 90.78\% (Data set 2, Java), and 55.73\% to 94.17\% (Data set 3, Java), but for the Java projects the accuracy suffers the most, as 11 out of 14 projects have less than 50\% microbenchmarks with at most a 1\% change rate across both data set 2 and 3.
All the \ac{RCIW}-based measures have a high accuracy, even for the Java projects, with \ac{RCIW}$_2$ having no project with less than 80\% microbenchmarks with a 1\% change in data set 2.
The \ac{RCIW}-based measures have, however, lower time savings, with 43.33\% to 86.19\% for the Go projects and 0\% to 50.11\% for the Java projects.
\ac{CV} has similar results to \ac{RCIW}$_1$, with \ac{CV} saving 76.8\% to 89.76\% (Go) and 3.21\% to 71.74\% (Java, across both data sets), and \ac{RCIW}$_1$ saving 65.85\% to 86.19\% (Go) and 0.68\% to 50.11\% (Java, across both data sets).
While the time savings from \ac{CV} are all higher, it has a lower result accuracy for the Java projects (the fraction of microbenchmarks with a 3\% and 5\% change rate is always lower for \ac{CV} than for \ac{RCIW}$_1$).

Comparing the results of the different stability metrics to both simple baselines, we observe that each strategy provides different ways to balance the trade-off between accuracy and time savings.
The minimum baseline has the highest possible time savings, but trades these for high accuracy loss.
The random baseline provides a better comparison point for the stability metrics, since it still provides high time savings, with still some accuracy loss notable in data set 2 and 3.
Out of the five stability metrics, RMAD shows results comparable to the random baseline, with RMAD reducing execution time even more with a higher accuracy loss mostly notable in data set 2 and 3.
RMAD performs favorably on data set 1, with time savings comparable to the minimum baseline, but providing higher accuracy.
The other metrics, CV, RCIW$_1$, RCIW$_2$, and RCIW$_3$, balance this trade-off better in the other direction, providing much higher accuracy for comparatively lower time savings.
We can therefore conclude, that the stability metrics provide a way to affect how µOpTime should balance time savings and accuracy loss, with accuracy results ranging from comparable to random (RMAD), to providing higher accuracy (all other metrics).

\summarybox{RQ1 Summary}{
    µOpTime reduces the execution time by 92.9\% to 95.97\% (Go) and 49.66\% to 94.17\% (Java), when using \ac{RMAD} as a stability metric.
    Other measures result in lower time savings, but higher accuracy.
    While the accuracy is always high for the Go projects, the Java projects often suffer from low accuracy.
    All the stability metrics except RMAD provide higher accuracy than the baselines, with RMAD still outperforming both baselines on data set 1.
}

\subsection{RQ2: How does the choice of the stability metric affect the execution time reduction?}
\label{sec:rq2}

\paragraph{Context}
In RQ2, we investigate the differences between the stability metrics to better understand their impact on time savings and accuracy.
As they all quantify stability in different ways, their impact also has to be considered in the context of different environments, such as the differences between Go and Java, as the JVM introduces more variability.

\paragraph{Methodology}
To check whether there are statistically significant differences between the results of µOpTime with different stability metrics, we use the Kruskal-Wallis~H test~\cite{Kruskal_Wallis_1952} to compare the distribution of change rates of each stability metric.
This test compares the medians of multiple samples, with the null hypothesis stating that all medians are the same.
If the null hypothesis can be rejected, we use Dunn's post-hoc test~\cite{Dunn_1964}, which compares all pairs to show, which have a statistically different median.
We use a significance level of $\alpha = 0.01$ with Bonferroni correction~\cite{shaffer_multiple_1995} where multiple comparisons are used.

For each statistically different pair of stability metrics, we also calculate Cliff's~$\delta$~\cite{cliff_1996} to quantify how strongly they differ.
If the observations of two groups are the same, then $\delta = 0$.
Should group 1 be larger than group 2, then $\delta > 0$, otherwise $\delta < 0$.
Cliff's~$\delta$ is bounded by $|\delta| \leq 1$.
We also use the following commonly used categories~\cite{romano_kromrey_2006}:
negligible ($|\delta| < 0.147$), small ($0.147 \leq |\delta| < 0.33$), medium ($0.33 \leq |\delta| < 0.474$), large ($|\delta| \geq 0.474$).

\paragraph{Results}
The Kruskal-Wallis~H test shows no statistically significant differences for all projects in data set 1 (Go) and data set 3 (Java), while all projects in data set 2 (Java) contain statistically significant differences between the stability metrics.
Thus, we use Dunn's post-hoc test on all projects in data set 2.

Dunn's post-hoc test reports a statistically significant difference between \ac{RMAD} and every other stability metric for all software projects, except for \emph{jenetics} and \emph{rxjava}.
For \emph{jenetics}, only \ac{RMAD} and \ac{RCIW}$_2$ are statistically different, while for \emph{rxjava} all pairs of stability metrics, except for \ac{RCIW}$_1$ and \ac{RCIW}$_3$, are statistically different.

For all the software projects, aside from \emph{jenetics} and \emph{rxjava}, $|\delta|$ ranges from ``small'' to ``large'' for all pairs involving \ac{RMAD}.
For \emph{jenetics}, the only statistically significant pair (\ac{RMAD} and \ac{RCIW}$_2$) shows a ``medium'' effect size.
For \emph{rxjava}, we investigate all pairs, except \ac{RCIW}$_1$ and \ac{RCIW}$_3$ further.
\ac{RCIW}$_2$ and \ac{RCIW}$_3$ have a ``negligible'' effect size, while all pairs involving \ac{RMAD} have a ``medium'' to ``large'' effect size.
All other pairs have a ``small'' effect size.

We conclude that the choice of stability metric largely does not matter for the software projects written in Go, as it has no statistically significant impact for all the software projects in data set 1.
For the Java projects, \ac{RMAD} produces consistently statistically different results, which, together with the overall lower accuracy (see \Cref{sec:rq1}), leads us to conclude that \ac{RMAD} leads to worse results.
In the case of \emph{rxjava}, even more differences are statistically significant, but only \ac{RMAD} has differences larger than ``small'' to the other stability metrics.
To further illustrate this point using the data from RQ1, using RMAD for \emph{rxjava} leads to 64.1\% of microbenchmarks to deviate more than 10\% from their original result.
Even for data set 3, where the Kruskal-Wallis~H test did not indicate statistically significant differences between stability metrics, using \ac{RMAD} still produces higher change rates in most projects.

Therefore, we recommend choosing either \ac{CV} or \ac{RMAD} for software projects written in Go, and \ac{RCIW}$_1$ or \ac{RCIW}$_3$ for software projects written in Java.
As \ac{CV} and \ac{RCIW}$_1$ are based on the mean, they should be used when using mean performance results, while \ac{RMAD} and \ac{RCIW}$_3$ should be used with median performance results, as they are based on the median.
For all the following sections, we choose \ac{RMAD} for the Go projects, and \ac{RCIW}$_3$ for the Java projects, to simplify further analysis.

The result data for this research question is available in our replication package.

\summarybox{RQ2 Summary}{
    While there are no statistical differences between the stability metrics for the Go projects, there are statistical differences for the Java projects.
    After considering the effect sizes using Cliff's $\delta$, we recommend choosing either \ac{CV} or \ac{RMAD} for Go projects, and \ac{RCIW}$_1$ or \ac{RCIW}$_3$ for Java projects for a minimal impact on accuracy.
}

\subsection{RQ3: How accurate is µOpTime when considering the warmup phase of Java microbenchmarks?}
\label{sec:rq3}

\begin{figure}[tbp]
	\centering
	\includegraphics[width=0.6\textwidth]{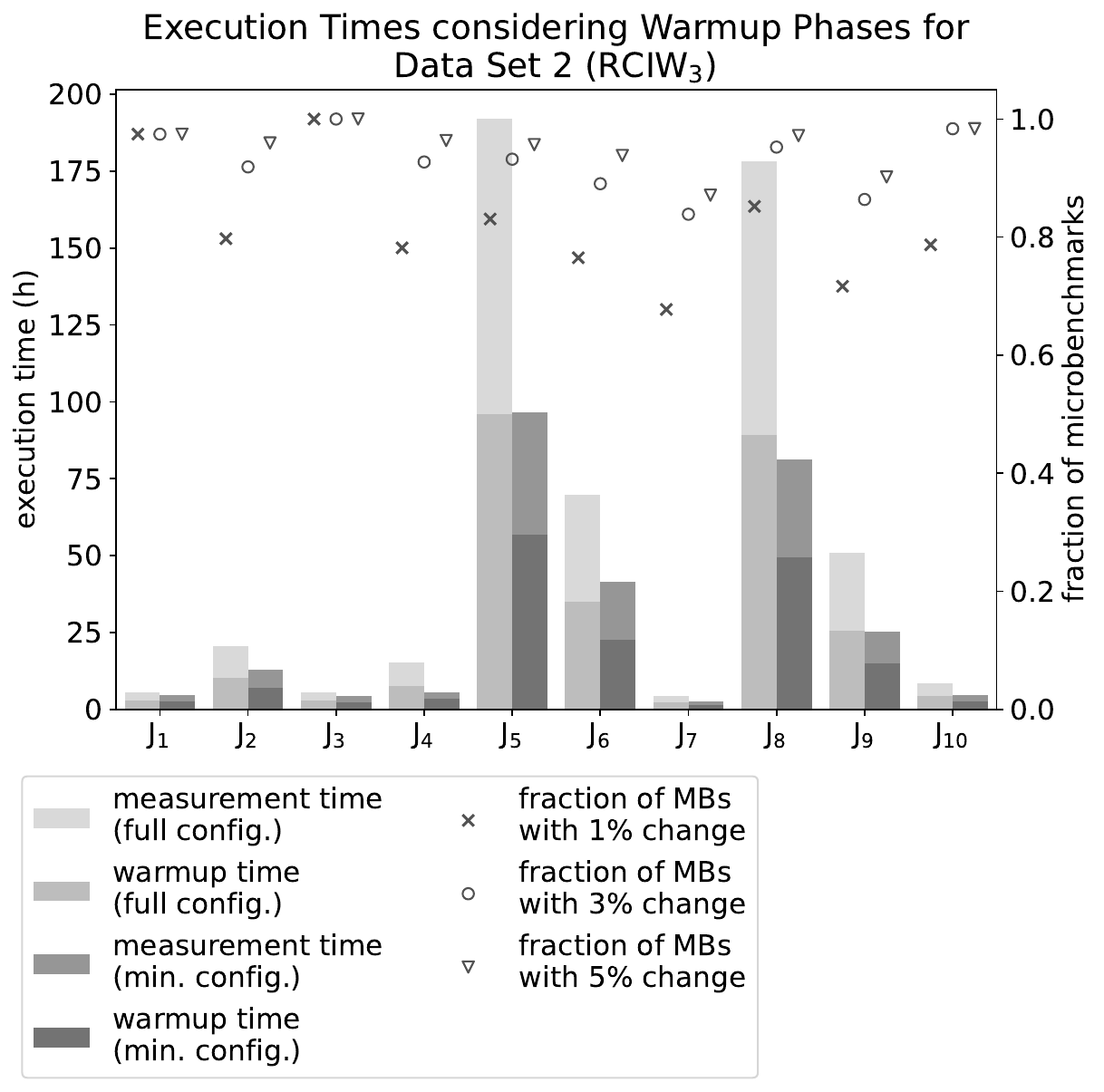}
    \caption{Time savings of µOpTime with warmup phase: The x-axis shows every project from data set 2.
    The left y-axis shows the execution time in hours of the bars, split into warmup and measurement phase.
    The right y-axis shows different fractions of microbenchmarks (MBs) with low change rates after using µOpTime, represented by marks in the figure (e.g., the fraction of MBs with a change rate smaller than 1\%).}
	\label{fig:rq3-time-savings-warmup}
\end{figure}

\paragraph{Context}
The lower performance values for the Java projects, especially regarding inaccuracy, may be due to the warmup phases of the microbenchmarks.
Due to the JVM's dynamic compilation, the performance can change depending on how long a program runs.
\citet{laaber_dynamically_2020} have studied how warmup phases can be shortened without impacting result quality,
but µOpTime does not consider a dedicated warmup phase, because its goal is to optimize measurement iterations.
In this RQ, we study the impact of optimizing measurement iterations with µOpTime in the presence of warmup iterations.

\paragraph{Methodology}
As we investigate the impact of JVM warmup behavior, we only use the Java projects from data set 2 in RQ3.
Data set 1 and 3 are not suitable for this analysis, since data set 1 contains Go microbenchmarks, which do not require warmup iterations, and data set 3 already has warmup iterations removed from the data set.
We assign a fixed number of warmup iterations to each fork, which are discarded from the result.
After this \emph{static warmup phase}, µOpTime minimizes the subsequent \emph{measurement phase}.
For our experiment, the static warmup phase consists of 50 warmup iterations, followed by 50 measurement iterations in the full execution configuration.
This setup was used by \citet{laaber_dynamically_2020} as a baseline, as it matches the JMH defaults for the warmup phase.
Although we use the same data set as \citet{laaber_dynamically_rp_2020}, we do not apply their approach to minimizing warmup phases, called \emph{dynamic reconfiguration}, since it would interfere with µOpTime.
Dynamic reconfiguration can stop new forks from being executed, which are also part of what µOpTime minimizes.
Furthermore, since the warmup phase runs on each fork of the JVM, warmup time can still get shortened whenever µOpTime minimizes the number of forks.
Nevertheless, both approaches minimize different parts of the benchmarking process and, therefore, could potentially be combined, which we consider subject of future work.
Here, we neither use nor compare with dynamic reconfiguration, as it has a different objective than µOpTime.

\paragraph{Results}
\Cref{fig:rq3-time-savings-warmup} summarizes the performance of µOpTime using \ac{RCIW}$_3$ with such a warmup phase.
Counter-intuitively, time savings are considerably higher than without using a warmup phase, ranging from 12.14\% to 63.58\% of overall execution time, with median time savings of 42.77\%.
Accuracy is high, with all projects having more than 80\% of microbenchmarks with a change rate smaller than 3\% (median of 92.96\%).
This behavior is explained by the warmup behavior of the JVM.
The JVM dynamically optimizes running code, leading to worse performance in the beginning, which then steadily becomes better, until it reaches a steady state~\cite{traini_towards_2023}.
When we did not consider warmup, even iterations, which had not reached the steady state yet, were part of the measurement phase, increasing the overall instability of the microbenchmark (i.e., increasing the chosen stability metric).
As these iterations are now part of the discarded warmup iterations, the instability is much lower, since only steady state measurements are used for calculating the stability.
This leads to µOpTime finding smaller execution configurations, even though the number of warmup iterations per fork remain fixed.
Therefore, we conclude that µOpTime performs better when considering warmup phases.

\summarybox{RQ3 Summary}{
    µOpTime performs considerably better on the Java projects, when warmup behavior is factored in, leading to time savings ranging from 12.14\% to 63.58\% of the overall execution time, with high accuracy as all the studied Java projects have more than 80\% microbenchmarks with a change rate smaller than 3\%.
}

\subsection{RQ4: How effective are the reduced execution configurations in detecting performance changes in CI/CD pipelines?}
\label{sec:rq4}

\begin{table}[tbp]
    \caption{Detected performance changes by the full and minimal execution configurations in Go projects using the minimum baseline, random baseline, and µOpTime.
    We sum up all detected performance changes by the full and minimal configuration over all studied version pairs.
    We outline the range of FPR and FNR across version pairs (=$v$ pairs), as well as the median FPR and FNR in parentheses.
    A detailed overview of the specific versions we studied can be found in the online appendix in our replication package.}
    \label{tab:res-perf-changes-go}
    \centering
    \begin{tabular}{lllrrrrr}
        \toprule
        Project & \#{}$v$ pairs & \#{}MB & Method & \multicolumn{2}{c}{\#{}Perf.\ Changes} & FPR & FNR \\
        \cmidrule{5-6}
        & & & & full & min & & \\
		\midrule
        \multirow{3}{*}{\emph{prometheus/common}} & \multirow{3}{*}{4} & \multirow{3}{*}{31} & minimum & 28 & 27 & 0\% (0\%) & 0\% -- 3.57\% (0\%)\\
        & & & random & 28 & 27 & 0\% (0\%) & 0\% -- 3.57\% (0\%)\\
        & & & µOpTime & 28 & 28 & 0\% (0\%) & 0\% (0\%)\\
        \midrule
        \multirow{3}{*}{\emph{pelletier/go-toml}} & \multirow{3}{*}{4} & \multirow{3}{*}{35} & minimum & 26 & 25 & 0\% (0\%) & 0\% -- 8.33\% (0\%)\\
        & & & random & 26 & 26 & 0\% (0\%) & 0\% (0\%)\\
        & & & µOpTime & 26 & 26 & 0\% -- 4.35\% (2.17\%) & 0\% -- 8.33\% (4.17\%)\\
        \midrule
        \multirow{3}{*}{\emph{uber-go/zap}} & \multirow{3}{*}{4} & \multirow{3}{*}{21} & minimum & 3 & 4 & 0\% -- 10\% (0\%) & 0\% -- 50\% (0\%)\\
        & & & random & 3 & 4 & 0\% -- 5\% (0\%) & 0\% (0\%)\\
        & & & µOpTime & 3 & 4 & 0\% -- 5\% (0\%) & 0\% (0\%)\\
        \bottomrule
    \end{tabular}
\end{table}

\begin{table}[tbp]
    \caption{Detected performance changes by the full and minimal execution configurations in Java projects using the minimum baseline, random baseline, and µOpTime.
    We sum up all detected performance changes by the full and minimal configuration over all studied version pairs.
    We outline the range of FPR and FNR across version pairs (=$v$ pairs), as well as the median FPR and FNR in parentheses.
    A detailed overview of the specific versions we studied can be found in the online appendix in our replication package.}
    \label{tab:res-perf-changes-java}
    \centering
    \begin{tabular}{lllrrrrr}
        \toprule
        Project & \#{}$v$ pairs & \#{}MB & Method & \multicolumn{2}{c}{\#{}Perf.\ Changes} & FPR & FNR \\
        \cmidrule{5-6}
        & & & & full & min & & \\
		\midrule
        \multirow{3}{*}{\emph{raphw/byte-buddy}} & \multirow{3}{*}{30} & \multirow{3}{*}{18 -- 20} & minimum & 11 & 48 & 0\% -- 25\% (5.56\%) & 0\% -- 100\% (0\%)\\
        & & & random & 11 & 7 & 0\% -- 10\% (0\%) & 0\% -- 100\% (0\%)\\
        & & & µOpTime & 11 & 28 & 0\% -- 10\% (0\%) & 0\% -- 66.67\% (0\%)\\
        \midrule
        \multirow{3}{*}{\emph{jenetics/jenetics}} & \multirow{3}{*}{11} & \multirow{3}{*}{34 -- 39} & minimum & 8 & 39 & 0\% -- 15.63\% (2.94\%) & 0\% -- 100\% (0\%)\\
        & & & random & 8 & 15 & 0\% -- 9.38\% (0\%) & 0\% -- 100\% (0\%)\\
        & & & µOpTime & 8 & 11 & 0\% -- 3.03\% (0\%) & 0\% (0\%)\\
        \midrule
        \multirow{3}{*}{\emph{JetBrains/xodus}} & \multirow{3}{*}{9} & \multirow{3}{*}{47} & minimum & 20 & 48 & 4.44\% -- 17.39\% (9.30\%) & 0\% -- 100\% (75\%)\\
        & & & random & 20 & 31 & 0\% -- 12.77\% (4.44\%) & 0\% -- 100\% (50\%)\\
        & & & µOpTime & 20 & 28 & 0\% -- 6.98\% (2.13\%) & 0\% (0\%)\\
        \midrule
        \multirow{3}{*}{\emph{openzipkin/zipkin}} & \multirow{3}{*}{9} & \multirow{3}{*}{49 -- 52} & minimum & 49 & 78 & 0\% -- 16\% (0\%) & 0\% -- 100\% (0\%)\\
        & & & random & 49 & 47 & 0\% -- 11.11\% (0\%) & 0\% -- 100\% (0\%)\\
        & & & µOpTime & 49 & 57 & 0\% -- 6.38\% (0\%) & 0\% -- 100\% (0\%)\\
        \bottomrule
    \end{tabular}
\end{table}

\paragraph{Context}
Finally, we embed µOpTime in a realistic CI/CD pipeline to showcase its applicability inside a real-world software development cycle.
This determines, whether µOpTime is a viable approach toward faster execution of microbenchmarks in an automated setting.

\paragraph{Methodology}
We only use the Go and Java projects of data set 1 and 3, as data set 2 does not contain multiple software versions.

To detect a performance change from a version $v_1$ to the next version $v_2$, we first run all the microbenchmarks using the full \ac{RMIT} configuration, which is $(3,3,5,5)$ for data set 1 and $(1,3,20,1)$ for data set 3, collect performance measurements, and calculate \ac{CI}s for both versions (percentile interval, $\alpha = 0.01$).
Then, we compare the microbenchmarks that exist in both versions.
To detect performance changes, we use non-overlapping \acp{CI} for both versions to indicate a performance change from $v_1$ to $v_2$, as also done in previous research~\citep{GrambowUMBS,laaber_applying_2021}.
We further define a \emph{relevant} performance change as a performance change of at least 3\%, to not consider changes due to measurement noise~\citep{georges_statistically_2007, huang_performance_2014}.

We investigate to what extent the minimal configuration from µOpTime, minimum baseline, and random baseline (introduced in \Cref{sec:rq1}) detect the same performance changes as the full configuration.
For all projects in data set 1 and 3, we compare all available versions in sequence, i.e., we compare $v_1$ to $v_2$, $v_2$ to $v_3$, and so on.
The available versions are listed in \Cref{tab:datasets} in \Cref{subsec:dataset}.
For some projects, we were unable to compare microbenchmarks beyond a certain version, because the developers replaced all microbenchmarks, which gives us a smaller number of usable version pairs for some projects.
Our ground truth are the relevant performance changes detected by the full configuration.
For each microbenchmark, we note whether the minimal configuration reports a true/false positive/negative, and calculate the \emph{false positive rate} (FPR) and \emph{false negative rate} (FNR).
Since we always compare µOpTime and the two baselines to the full execution using the same underlying software versions and data, any code changes affect both the minimal and full execution.

\paragraph{Results}
\Cref{tab:res-perf-changes-go} shows the results for data set 1, and \Cref{tab:res-perf-changes-java} shows the results for data set 3.
The minimized configurations from µOpTime perform exceptionally in detecting the relevant performance changes in most cases.

For the Go projects, the minimized configurations using µOpTime identify 55 out of 57 performance changes correctly, and wrongly detect or miss performance changes for just 3 out of 12 version pairs.
The FPR never exceeds 5\%, while the FNR is at most 8.33\%, with just 3 false positives and 2 false negatives in 348 microbenchmark comparisons.
The baselines perform similarly well, with the minimum baseline having the same number of false positives and negatives, just at different microbenchmarks.
The random baseline performs the best, with only 1 false positive and 1 false negative.
Considering the time savings reported in \Cref{sec:rq1}, this is not too surprising, as the stability metric we use here, RMAD, has almost the same time savings as the minimum baseline, while the random baseline takes more time.
The difference in quality inside a CI/CD pipeline is almost negligible as well, therefore we conclude that µOpTime works for Go projects, although our baselines perform similarly well.

For the Java projects, the minimized configurations using µOpTime identify 82 out of 89 performance changes correctly, and wrongly detect or miss performance changes for 25 out of 59 version pairs.
The FPR never exceeds 10\%, while the FNR can reach 100\%.
This is due to a low number of true positives (full configuration performance changes), where missing just one performance change already highly increases the FNR, such as for \texttt{v2.5.0} and \texttt{v2.6.0} of zipkin, where missing one performance change increases the FNR to 100\%.
Overall, there are 29 false positives and 7 false negatives in 1,844 microbenchmark comparisons.
Here, the baselines both perform worse than using µOpTime.
The minimum baseline reports 166 false positives and 45 false negatives over the same microbenchmarks and performance changes, while the random baseline fares slightly better with 53 false positives and 41 false negatives overall.
This shows that for the Java projects, µOpTime achieve considerably better results than the baseline approaches.

Therefore, we can conclude that µOpTime works well in CI/CD pipelines for both the Go and Java projects.
This represents a first step toward making large microbenchmark suites viable for automated performance testing.

\summarybox{RQ4 Summary}{
    The microbenchmark suites with minimal execution configurations from µOpTime detect 55 (82) performance changes out of the 57 (89) detected by the full execution configuration in Go (Java) projects.
    The quantity of false positives (3 for Go, 29 for Java) and false negatives (2 for Go, 7 for Java) are low (overall microbenchmark comparisons are 348 for Go, 1,844 for Java), indicating that the minimal execution configurations do not miss or wrongly detect performance changes, compared to the full execution configuration.
}

\section{Discussion}
Optimizing microbenchmark suites using individual execution configurations reduces the total suite execution time drastically while still ensuring relevant measurements.
Avoiding unnecessary microbenchmark executions, microbenchmark suites can fit the requirements of modern cloud-based development cycles and can be embedded in regular CI/CD pipelines requiring a fast performance feedback.
Applying our approach, however, also includes several important factors to keep in mind which we discuss in the following.

\paragraph{Managing detected performance changes}

Committed code changes fix issues or implement new features into the software, but can also introduce major performance problems as a side effect. 
A good CI/CD pipeline should detect them and reject the corresponding change, or ask for confirmation from developers.

If there is a performance issue detected in a microbenchmark, it is important to understand how relevant it is to the application. 
While functionalities such as login, save, and load are probably highly relevant functionalities that are executed frequently and cannot tolerate significant performance regressions, there are certainly others such as a monthly data export or rare business transactions that tolerate greater margins.
Thus, the performance change detection threshold must be adjusted to each microbenchmark individually or can be a generic one complying with the hardest requirements.

Furthermore, some changes, such as the roll-out of more complex security functionalities, imply a slowdown.
For example, using a more complex encryption algorithm is likely to slow down every interaction but is intended and should not be rejected by the pipeline.
Thus, a confirmation mechanism can be used to ignore or overwrite detected performance drops if the change is either related to rarely used functionalities or was expected.

\paragraph{Estimation of FPR and FNR}

We estimate the FPR and FNR for µOpTime by assuming the full execution provides the ground truth.
The full execution can, however, also measure false positives or false negatives compared to the actual \enquote{true} performance regressions.
As we do not have perfect insight into the \enquote{true} performance regressions of real world software, using the full execution as the comparison is the best approximation to the actual FPR and FNR, which we can provide.
This is weighted in favor of the full execution, as µOpTime can only perform worse using this method of comparison.
More sophisticated approaches, such as techniques used in FBDetect~\cite{fbdetect}, could identify false positives, which are caused by outside influence, such as cloud performance variability, and estimate the FPR and FNR closer to the underlying truth.

\paragraph{Influence of inherent stability differences between Go and Java}

The results of how well µOpTime reduces the execution configuration in Go and Java microbenchmarks show that the Go microbenchmarks we studied are generally more stable than the Java microbenchmarks.
This might be due to a lot of different factors, such as Go being a language compiled into native binaries, while Java programs still need to run on the JVM, which introduces performance variability.
This leads to µOpTime performing similarly to the baseline approaches for Go, while it outperforms them for Java, as the trade-off between accuracy and time savings does not matter as much for Go.
Because of such influences, our results are strictly limited to Go, Java, and the specific software projects we studied.
Any other programming language and software project might have different factors introducing more (or less) inherent performance variability, making µOpTime more relevant (or less relevant) for that particular language or project.

\paragraph{Influence of load changes in the cloud on measurement results}

Load changes during an experiment can affect microbenchmarking results, and therefore the minimized execution configurations, which µOpTime produces, as well as how well these perform in practice.
This does not just apply to short-term performance fluctuations, but also to mid- to long-term performance fluctuations, such as weekly patterns.
This problem applies to data set 1 in particular, as data set 2 and 3 were collected in a controlled bare-metal environment.
The previously mentioned FBDetect~\cite{fbdetect} can also identify false positives caused by load changes in the cloud and mitigate this problem.

\paragraph{The optimization must be repeated after major changes}

The evaluated software, its microbenchmarks, and its execution environment can gradually change over time. 
For example, new functionalities are covered by additional microbenchmarks, the workload of microbenchmarks may change due to new requirements, new hardware speeds up some functionalities, or an operating system update slows down some.
While a new benchmark needs to be optimized anyway, a change either in the workload definition, in the evaluated code sections, or the execution environment can affect the stability of microbenchmarks.
Similarly, changing the repetition levels of the full execution configuration will likely lead to different results for some microbenchmarks, particularly if they were not optimized before.
Thus, the optimization must be repeated, e.g., for every major version or whenever necessary, to adjust the individual execution configurations of each microbenchmark to the current code version.
How often the optimization needs to be repeated depends on many factors unique to each software project.
Therefore, we cannot give general guidelines on how often the optimization needs to be performed in practice.
In practice, the full execution and optimization could be run \emph{offline}, while the optimized execution is run \emph{online} directly after a commit, to maximize time savings in CI/CD pipelines, and still repeat the optimization process.
Our results from RQ4 indicate that using the minimal execution for several versions worked well for the studied projects.

\paragraph{Ensure proper RMIT execution}

With our optimization, stable microbenchmarks require fewer repetitions and will stop earlier, e.g., after the second suite run. 
Thus, fewer and fewer microbenchmarks will be executed with each suite run.
In an extreme case with many stable microbenchmarks and a single unstable one, this would end up in several suite runs with only a single microbenchmark left which somewhat invalidates the idea of RMIT and might affect the results.
To overcome this issue, we recommend tooling which detects these cases and ensures a minimal number of different microbenchmarks in each repetition.

\paragraph{Full execution configuration uses insufficient measurement iterations}

In our experiments and also in the data sets used, each microbenchmark was repeated many times to ensure reliable results. 
For single specific microbenchmarks, however, this full execution configuration might not repeat the microbenchmark \enquote{often enough} to yield reliable results.
This might be because it would require even more suite repetitions or iterations, making the suite execution time even longer and, therefore, infeasible to execute.
We see this especially when applying µOpTime:
Ideally, running the full execution configuration should repeat each microbenchmark more often than necessary to ensure relevant results. 
With our optimization, there should then be a reduction in each microbenchmark, which is not always the case. 
In the ten Java projects of data set 2 we examined, the execution configurations of 1,414 out of 3,969 microbenchmarks (35.63\%) were not reduced, when RCIW$_3$ was used as a stability metric, and warmup is considered.
Thus, stability measure and configuration must be aligned, and in doubt, more repetitions must be added to the full execution configuration.
Additionally, as the stability of a microbenchmark might change over time, a full execution configuration might become insufficient over time.
This problem can only be caught, if the full execution is repeated.
We evaluated, how well µOpTime performs in CI/CD pipeline in \Cref{sec:rq4} compared to the full execution, but we do not consider the problem of the full execution becoming insufficient itself.
When collecting data set 1, we increased the parameter \emph{seconds per iteration} from 1 second in previous work to 5 seconds for our data collection to mitigate this problem, as the experiments ran in a cloud environment.

\paragraph{Integration with alternative execution time reduction approaches}

As µOpTime is only concerned with optimizing the measurement phase of microbenchmarks, it is orthogonal to other approaches, which optimize warmup time~\cite{laaber_dynamically_2020}, or reduce the microbenchmark suite by selecting a subset of microbenchmarks to run for a given experiment~\cite{GrambowUMBS, Grambow2021}.
As such, these approaches have potential to be combined for higher time savings.
A combined setup of these methodologies needs to be critically evaluated in itself, however, since more complex interplay between these methods might affect result quality more than each individual method does.

\paragraph{Influence of microbenchmark coverage on µOpTime}

In this paper, we do not investigate influences of microbenchmark coverage on µOpTime.
Since we use open-source software and public data sets to evaluate µOpTime, and crucially only include microbenchmarks added by the original developers, we have no control over microbenchmark coverage.
Since µOpTime optimizes \emph{individual} microbenchmarks, and does not take other information of the code into account, it should theoretically work with any number of microbenchmarks regardless of coverage.
Should a project have only a very small number of microbenchmarks, and therefore low coverage, we run into the problem that proper RMIT execution cannot be guaranteed.

\section{Threats to Validity}

\paragraph{Construct Validity}
The main threats to construct validity relate to the metrics used to compute the stability and time savings as well as the effectiveness in detecting changes.

We use the change rate $c$ (RQ1, RQ2, and RQ3) of the benchmark's optimized and full configuration and use three thresholds (1\%, 3\%, and 5\%) to decide whether a benchmark is accurate or not.
Different metrics to assess the change and thresholds are likely to lead to different results and conclusions.
However, we base both the change rate and thresholds on previous performance engineering research~\citep{georges_statistically_2007,huang_performance_2014}.

In terms of time savings (RQ1, RQ2, and RQ3), we rely on the benchmark suite execution time difference between the suite with the full configurations and the one with the optimized configurations.
Different measures of time savings, such as per-benchmark time savings, would probably lead to different results.
However, as the suite time savings are based on the benchmark time savings, we consider this threat negligible.

For performance changes between two consecutive versions (RQ4), we use non-overlapping bootstrap (percentile) CIs, similar to previous research~\citep{bulej:17a,laaber_dynamically_2020,he_statistics_2019}.
Different performance change detection techniques can lead to different results.
In addition, we use FPR and FNR to show whether the optimized benchmark configurations are able to comprehensively detect real performance changes.

\paragraph{Internal Validity}
The threats to internal validity mostly relate to the performance measurements and execution platforms.
For data set 1, we employ (traditionally unreliable) cloud infrastructure, which can have a drastic impact on the measured performance, as load changes due to background tasks, parallel jobs, co-located tenants, or other reasons impact the cloud performance~\citep{LeitnerCito2016}.
We follow best practice in performance measurements in the cloud~\citep{LaaberScheunerLeitner2019, papadopoulos:19} to reduce this threat.
Moreover, we use repeated iterations and apply the RMIT methodology~\citep{AbediCloud} to randomize execution orders to lower the impact of periodically-occurring confounding factors.
In addition, we rely on two data sets (2 and 3) from previous research that use bare-metal machines as execution environment to reduce the cloud threat for the overall results.

In RQ4, we use the measured performance changes between two versions as the ground truth.
These changes could be due to measurement inaccuracies or environment noise.
To mitigate this threat, we follow a rigorous measurement methodology (as also discussed above) and only consider changes above a threshold of 3\%.
The chosen threshold is based on previous research~\citep{georges_statistically_2007,huang_performance_2014}.
Different ways to identify changes, e.g., from the issue tracker, or thresholds would potentially change our results.

\paragraph{External Validity}
Generalizability of our results is mostly concerned with the studied projects and execution platforms.

Our study investigates 14 open-source projects written in Go and Java.
Consequently, we cannot claim that our results will hold for other projects and languages, or commercial software.
As performance experiments are time-intensive, a more extensive experimental setup quickly becomes infeasible.
We aim at representative projects of different application domains to improve generalizability.

In terms of execution platforms, we study benchmark results from both bare-metal and cloud environments to increase external validity.
However, different environments might lead to different stability results and, consequently, different conclusions.

\section{Related Work}

µOpTime is related to a number of other approaches for reducing benchmark execution time. 
To the best of our knowledge, we are the first proposing individual execution configurations for each microbenchmark using a static, offline approach and exploring the whole parameter space of measurement iterations.
Our approach can be used in CI/CD pipelines, potentially running on cloud infrastructure, to detect performance issues as they arise, which is constantly gaining in importance and relevance. 
While we use a solid median-based comparison mechanism to quantify performance differences, various others can be used equally well, depending on the application scenario.

\paragraph{Reducing Benchmark Execution Time}
\label{sec:stopbench_relwork}

Benchmarks lasting several hours usually cannot be integrated into regular CI/CD pipelines as they prevent fast performance feedback and massively slow down the whole process.
Thus, reducing the execution time without losing relevant insights is an important issue in research. 

First, several approaches rely on stopping live benchmarking experiments as soon as measurements are stable~\cite{he_statistics_2019, AlGhamdi2016, AlGhamdi2020, laaber_dynamically_2020, traini_ai-driven_2024, metior}.
Using statistical methods such as the Kullback-Leibler divergence or the Wilcoxon test, benchmarks are iterated until no difference to previous measurements can be detected.
While this comes with the advantage that there is no need to specify an individual execution configuration for each microbenchmark at all, the optimization can only focus on \textit{one} predefined dimension: 
shortening the benchmark duration, number of iterations, number of suite runs, \textit{or} number of cloud instances.
Using µOpTime, however, the offline analysis can evaluate all possible configuration combinations to determine the best one.

Other approaches for shortening the duration focus on the targeted execution of specific microbenchmarks~\cite{Grambow2021, javed_test_2022, chen_perfjit_2020, de_oliveira_perphecy_2017}.
Here, a subset of microbenchmarks is selected by criteria such as practical relevance, code coverage, or past measurements. 
All these approaches can be applied separately and combined with µOpTime to further reduce the benchmark duration.

A third approach is to apply test case prioritization techniques to the benchmark area~\cite{rothermel_test_1999, mostafa_perfranker_2017, laaber_applying_2021, laaber_mobp_2022}.
Running critical and relevant microbenchmarks first that already detect any degradation in the first iterations makes the execution of further benchmarks unnecessary and delays the result.
As µOpTime uses RMIT execution of microbenchmark suites and always explores the whole parameter space of measurements, a prioritization of individual microbenchmarks only makes sense when using the reduced execution configurations in regular CI/CD pipelines;
however, finding ways to leverage prioritization and µOpTime together is desirable and subject to future research.

Finally, static source code features can also be used to predict unstable microbenchmarks and adjust the execution configuration accordingly~\cite{laaber_predicting_2021}.
While this can be used to estimate a good individual execution configuration without actually executing the microbenchmarks, µOpTime determines the configuration by first executing the full configuration and identifying the best minimal configuration in a subsequent offline analysis.

\paragraph{Benchmarking in CI/CD pipelines}
\label{sec:cicd_relwork}

Besides our study, there are several others that integrate microbenchmarks into CI/CD pipelines to detect performance changes~\cite{javed_perfci_2020, waller_including_2015, GrambowUMBS, daly_industry_2019, daly_creating_2021, LaaberMBEval}.
Application benchmarks, i.e., stressing fully set-up systems such as a database system with an artificial load such as HTTP requests~\cite{DavidCSB}, are used more and more for detecting performance issues~\cite{foo_mining_2010, ingo_automated_2020, GrambowLehmannBermbach2019, foo_industrial_2015}.
Furthermore, there is tooling to support cloud-based benchmark experiments~\cite{silva_cloudbench_2013, hasenburg_mockfog_2021,paper_pfandzelter_celestial}.

µOpTime is designed to be integrated into cloud-based CI/CD pipelines as well but focuses on microbenchmark suites only.

\paragraph{Detecting Performance Changes}
\label{sec:perfchange_relwork}

µOpTime reports changes when performance confidence intervals of two successive versions do not overlap.
There are, however, several other possible techniques for detecting and quantifying performance changes based on thresholds~\cite{GrambowUMBS, GrambowLehmannBermbach2019, foo_mining_2010}, machine learning~\cite{iter8}, or others~\cite{LaaberMBEval, daly_industry_2019, matteson_nonparametric_2014, fbdetect}.
The performance change detection and quantification algorithms of µOpTime can be replaced by any other, as long as they report both the performance value and respective confidence intervals.

\section{Conclusion}
In this paper, we introduced µOpTime as a static, offline approach to minimize the execution time of microbenchmark configurations with multiple levels of repetition, such as RMIT-based configurations, by reducing the number of measurements.
We showed that µOpTime can significantly reduce the execution time, reaching up to $95.83\%$ (median of $76.8\%$, across stability metrics) on software projects written in Go, and up to $63.58\%$ (median of $42.77\%$, using RCIW$_2$) on software projects written in Java, when a separate warmup phase is considered.
Finally, we showed that µOpTime performs well in CI/CD pipelines, detecting the majority of the performance issues in our study that a full configuration detects, with an occurrence of false positives of 3 (29) and false negatives of 2 (7) within 348 (1,844) microbenchmark comparisons for Go (Java) projects.

\begin{acks}
    C.~Laaber is supported by The Research Council of Norway project \texttt{309642}.
\end{acks}

\bibliographystyle{ACM-Reference-Format}
\bibliography{refs}

\end{document}